\documentclass[10pt]{iopart}
\usepackage{iopams,color}
\usepackage[colorlinks=true,linkcolor=blue,citecolor=red,bookmarksnumbered=true]{hyperref}
\usepackage{amssymb,times,fourier,charter,graphicx,subfigure}
\usepackage{fullpage}
\graphicspath{{figs}}

\advance\textwidth -16mm
\advance\hoffset 8mm
\advance\textheight -8mm
\advance\voffset 4mm

\eqnobysec

\graphicspath{ {./figs/} }
\makeatletter
\def\eqref#1{(\ref{#1})}
\def\note[#1]{\marginpar{{\color{red}[#1]}}}

\def\numparts{\refstepcounter{equation}%
     \setcounter{eqnval}{\value{equation}}%
     \setcounter{equation}{0}%
     \def\theequation{\arabic{section}.\arabic{eqnval}{\it\alph{equation}}}}
\def\endnumparts{\def\theequation{\arabic{section}.\arabic{equation}}%
     \setcounter{equation}{\value{eqnval}}}
\def\bse{\numparts}
\def\ese{\endnumparts}
\def\bea{\begin{eqnarray}}
\def\eea{\end{eqnarray}}
\def\be{\begin{equation}}
\def\ee{\end{equation}}
\definecolor{deeppurple}{rgb}{0.6,0,0.6}

\makeatletter
\definecolor{darkblue}{rgb}{0,0,0.8}
\renewcommand\section{\@startsection {section}{1}{\z@}%
  {-3.2ex \@plus -1ex \@minus -.2ex}%
  {2ex \@plus.2ex}%
  {\color{darkblue}\reset@font\normalsize\bfseries\raggedright}}
\renewcommand\subsection{\@startsection{subsection}{2}{\z@}%
  {-3ex\@plus -1ex \@minus -.2ex}%
  {1ex \@plus .2ex}%
  {\color{darkblue}\reset@font\normalsize\itshape\raggedright}}

\def\K{K_m}
\def\E{E_m}
\newcommand\Dy[1][]{\frac{D#1}{Dy}}
\let\grad=\nabla
\newcommand\deriv[3][]{\frac{\d^{#1}#2}{\d{#3}^{#1}}}
\newcommand\partialderiv[3][]{\frac{\partial^{#1}#2}{\partial {#3}^{#1}}}
\let\true@epsilon=\epsilon
\let\epsilon=\varepsilon
\let\true@phi=\phi
\let\phi=\varphi
\def\I{{\mathbb I}}

\def\Real{{\mathbb{R}}}

\def\Wr{\mathop{\rm Wr}\nolimits}
\def\diag{\mathop{\rm diag}\nolimits}

\def\cn{\mathop{\rm cn}\nolimits}
\def\sgn{\mathop{\rm sgn}\nolimits}

\def\d{{\rm d}}
\def\e{{\rm e}}

\def\@#1{{\mathbf{#1}}}
\let\truetilde=\~
\def\~#1{{\ifmmode\tilde#1\else\truetilde#1\fi}}

\def\_#1{{\mathsf{#1}}}
\def\txtfrac#1#2{{\textstyle\frac{#1}{#2}}}
\let\truenabla=\nabla
\def\fbf#1{\setbox0=\hbox{$#1$}\kern-0.10\wd0
  \lower0.02em\copy0\kern-\wd0 \lower0.02em\hbox{\kern+0.05em\copy0}\kern-\wd0
  \raise0.02em\copy0\kern-\wd0 \raise0.02em\hbox{\kern-0.05em\box0}}
\def\bfnabla{{\fbf\truenabla}}
\makeatother

\definecolor{alexgreen}{rgb}{0,0.4,0.4}

\newcommand{\eq}[1]{\begin{equation}#1\end{equation}}

\let\sub=\relax
\let\nsub=\relax

\let\dv=\deriv
\let\pdv=\partialderiv
\let\vb=\@
\let\qty =\relax

\let\grad = \bfnabla

\begin{document}

\title{\color{red}%
Two-dimensional reductions of the Whitham modulation system for the Kadomtsev-Petviashvili equation}
\author{Gino Biondini$^1$, Alexander J. Bivolcic$^1$, Mark A. Hoefer$^2$ and Antonio Moro$^3$}
\address{$^1$~State University of New York, Department of Mathematics, Buffalo, NY 14260, USA}
\address{$^2$~University of Colorado, Department of Applied Mathematics, Boulder, CO 80303, USA} 
\address{$^3$~Northumbria University, Department of Mathematics, Physics and Electrical Engineering, Newcastle, NE1 8ST, UK}
\date{\today}
\def\submitto#1{\vspace{28pt plus 10pt minus 18pt}
     \noindent{\small\rm To be submitted to : {\it #1}\par}}

\begin{abstract}
Two-dimensional reductions of the KP-Whitham system, namely the overdetermined Whitham modulation system for 
five dependent variables that describe the periodic solutions of the 
Kadomtsev-Petviashvili equation, are studied and characterized.
Three different reductions are considered corresponding to
modulations that are independent of $x$, independent of $y$, and of $t$ (i.e., stationary), respectively.
Each of these reductions still describes dynamic, two-dimensional spatial configurations
since the modulated cnoidal wave generically has a nonzero speed and
a nonzero slope in the $xy$ plane.
In all three of these reductions, the properties of the resulting systems of equations are studied.
It is shown that the resulting reduced system is not integrable unless one enforces 
the compatibility of the system with all conservation of waves equations 
(or considers a reduction to the harmonic or soliton limit).
In all cases, compatibility with conservation of waves 
yields a reduction in the number of dependent variables to two, three and four, respectively.
As a byproduct of the stationary case, the Whitham modulation system for the Boussinesq equation is also explicitly obtained.
\par
\kern\smallskipamount\noindent
\today
\end{abstract}

\section{Introduction and background} 
\label{s:background}

The description of dispersive wave propagation has been a classical
topic of study dating back to the works of Boussinesq, Stokes,
Rayleigh, Korteweg and de Vries and others in the nineteenth century,
and it continues to attract significant attention.  A scenario of both
theoretical and applicative interest is that in which dispersive
effects are much smaller than nonlinear ones, a regime that often
leads to the generation of dispersive shock waves.  Indeed, a large
number of works have been devoted to this subject (e.g., see
\cite{ElHoefer} and references therein). The mathematical framework
for the description of small dispersion problems in one spatial
dimension and the formation of dispersive shock waves in that context
have been well characterized, beginning with the seminal work of
G.~B. Whitham \cite{Whitham1974}.  However, our understanding of
dispersive wave propagation and dispersive shock waves in more than
one spatial dimension is much less developed.

The purpose of this work is to study special solutions of the
Kadomtsev-Petviashvili (KP) equation \cite{KP1970},
\be
(u_t+6uu_x+\epsilon^2 u_{xxx})_x + \sigma u_{yy} = 0\,,
\label{e:KPo}
\ee
where $0<\epsilon\ll1$, subscripts $x$, $y$ and $t$ denote partial
differentiation and the values $\sigma=\mp1$ distinguish between the
KPI and KPII variants of the KP equation, respectively.  The KP
equation, which is a two-dimensional generalization of the celebrated
Korteweg-de\,Vries (KdV) equation, similarly arises in such diverse
fields as plasma physics~\cite{InfeldRowlands,KP1970,Lonngren1998},
fluid dynamics~\cite{AS1981,Kodama2018}, nonlinear
optics~\cite{BWK16,PSK95} and ferromagnetic media~\cite{TF85}.
The KP equation is also, like the KdV equation, a completely
integrable infinite-dimensional Hamiltonian system whose solutions
possess a rich mathematical structure
\cite{AC1991,BBEIM1994,Hirota2004,InfeldRowlands,Kodama2018,Konopelchenko,NMPZ1984}.
The initial-value problem for the KP equation is in principle amenable
to exact solution via the inverse scattering transform (IST)
\cite{AC1991,Konopelchenko,NMPZ1984}.  Yet, even though considerable
work has been devoted to the development of the IST for the KP
equation throughout the last twenty years
\cite{IP17p937,JMP44p3309,TMP159p721,TMP165p1237,DerchyiWu2021}, the
IST has rarely been used to study the dynamical behavior of solutions
of the KP equation \cite{DerchyiWu2022}.
Conversely, asymptotic methods such as Whitham modulation theory have
recently been shown to be quite effective in this regard
\cite{Biondini_2017,Biondini_2020,NLTY21,JFM21}.

In this work, we derive and characterize several asymptotic reductions
of the KP equation, which we rewrite in evolution form as
\be
u_t +
6uu_x + \epsilon^2 u_{xxx}+\sigma v_y = 0\,,\qquad v_x = u_y\,.
\label{e:KP}
\ee
The linear dispersion relation of~\eqref{e:KP}, obtained by looking for small-amplitude plane-wave solutions 
$u(x,y,t) = u_o + A\, e^{i\theta(x,y,t)}$ with $|A| \ll |u_o|$,
$\theta(x,y,t) = (kx + ly -\omega t)/\epsilon$, 
is
$\omega = (6u_o + \sigma q^2)k - k^3$,
with $q= l/k$.
In addition, the KP equation admits nonlinear, exact traveling wave solutions in the form of ``cnoidal waves'' 
\be
u(x,y,t) = r_1-r_2 +r_3 +2(r_2-r_1)\cn^2(2\theta\K ; m)\,,\qquad
v(x,y,t) = qu+p\,,
\label{e:KP_cn_wave}
\ee
where $\cn(\cdot)$ denotes the Jacobian elliptic cosine \cite{NIST}, 
$K_m = K(m)$ and $E_m = E(m)$ are the complete elliptic integrals of the first and second kind, respectively,
and 
\be
\label{e:m_def}
m=\frac{r_2-r_1}{r_3-r_1}
\ee
is the elliptic parameter.  The above solution is completely
determined by five parameters: $r_1$, $r_2$, $r_3$, $q$ and~$p$.  The
local wavenumber $k$ and $l$ in the $x$ and $y$ directions and the
frequency $\omega$ are then obtained as 
\bse \bea k =
\label{e:klomegadef}
{\sqrt{r_3-r_1}}\,\big/{2\K}\,,
\label{e:def_k}
\\
l = qk\,,\\
\omega = (V + \sigma q^2)k\,,
\label{e:def_omega}
\eea
\ese
with
\be
V = \frac{\omega}{k}-\sigma q^2 = 2(r_1 + r_2 + r_3)\,.
\label{e:qV}
\ee

In \cite{Biondini_2017}, 
the method of multiple scales was used to derive the so-called KP-Whitham system, 
i.e., a system of quasilinear first-order PDEs that describes the slow modulation 
of the above periodic solutions of the KP equation.
One begins by seeking a solution of \eqref{e:KP} in the form $u=u(\theta,x,y,t)$, 
with rapidly varying variable $\theta(x,y,t)$ defined through its derivatives:
\be
\theta_x={k(x,y,t)}/{\epsilon}, \qquad 
\theta_y={l(x,y,t)}/{\epsilon}, \qquad 
\theta_t= - {\omega(x,y,t)}/{\epsilon}\,.
\ee
Here, $k(x,y,t)$ and $l(x,y,t)$ are the local wave numbers in the $x$ and $y$ directions, respectively,
and $\omega(x,y,t)$ is the wave's local frequency.
Imposing the equality of the mixed second derivatives of $\theta$ results in the compatibility conditions
\bse
\label{e:waveconservation}
\bea
k_t+\omega_x &=0\,,
\label{e:waveconservation1}\\
l_t+\omega_y &=0\,,
\label{e:waveconservation2}\\
k_y-l_x &=0\,,
\label{e:waveconservation3}
\eea 
\ese 
called the ``conservation of waves'' equations.  
One also
introduces the dependent variable \be q(x,y,t) = \frac{l}{k}\,
\label{e:qdef}
\ee
consistent with the above periodic solutions, along with the slowly varying variables $x$, $y$ and $t$.
It was then shown in \cite{Biondini_2017} that to leading order one recovers the solution~\eqref{e:KP_cn_wave}.
When the parameters of the above periodic solution are slowly modulated with respect to $x$, $y$ or $t$, 
they satisfy a system of Whitham modulation equations. 
When writing down these equations, it is convenient to define the ``convective derivative"
\vspace*{-0.6ex}
\be
\Dy{} = \partialderiv{}{y}-q\partialderiv{}{x}~.
\ee
In component form, the KP-Whitham system (KPWS) is then comprised of the following partial differential equations (PDEs) 
\bse
\label{e:KPWS}
\bea
\partialderiv{r_i}{t}+(V_j+\sigma q^2)\partialderiv{r_j}{x}+2\sigma q \Dy[r_j]+\sigma \nu_j\Dy[q]+\sigma\Dy[p]=0\,, \qquad j=1,2,3,
\label{e:KPWS-1}\\
\partialderiv{q}{t}+(V_2+\sigma q^2)\partialderiv{q}{x}+2\sigma q\Dy[q] + (4-\nu_4)\Dy[r_1] + (2+\nu_4)\Dy[r_3] =0\,,
\label{e:KPWS-2}\\
\partialderiv{p}{x} -
(1-\alpha)\Dy[r_1]-\alpha\Dy[r_3]+\nu_5\partialderiv{q}{x} = 0\,,
\label{e:KPWS-3}\\
b_1\Dy[r_1] + b_2\Dy[r_2] + b_3\Dy[r_3] + b_4\partialderiv qx = 0\,.
\label{e:qconstraint}
\eea
\ese
[Note that~\eqref{e:waveconservation1} is a consequence of the three equations~\eqref{e:KPWS-1},
while \eqref{e:waveconservation2} and~\eqref{e:waveconservation3} are equivalent to~\eqref{e:KPWS-2} and~\eqref{e:qconstraint},
respectively.]
Here,
$V_1,\dots,V_3$ are the characteristic speeds of the Whitham system for the KdV equation,
namely
\vspace*{0.4ex}
\bea
\label{e:Vdef}
V_1 = V - 2b \frac{\K}{\K - \E}\,,\quad
V_2 = V - 2 b \frac{(1-m) \K}{\E - (1-m) \K}\,,\quad 
V_3 = V  + 2b \frac{(1-m) \K}{m\E}\,,
\nonumber\\
\eea
and
$b = 2(r_2-r_1)$ is the amplitude of the cnoidal wave solution~\eqref{e:KP_cn_wave},
while the remaining coefficients are 
\bse
\label{e:nudef}
\bea
\nu_1 = \frac{V}{6} + \frac{b}{3 m} \frac{(1+m)\E-\K}{\K-\E}\,, \quad
\nu_2 = \frac{V}{6} + \frac{b}{3 m} \frac{(1-m)^2\K - (1-2 m)\E}{\E - (1-m)\K}\,,\\
\nu_3 = \frac{V}{6} + \frac{b}{3 m} \frac{(2-m)\E -(1-m)\K}{\E}\,, \quad
\nu_4 = \frac{2m\E}{\E - (1-m) \K}\,,\\
\nu_5 = r_1 - r_2 + r_3\,,\quad
\alpha = \frac{\E}{\K}\,,\quad b_1 = (1-m)(\K-\E),\\ b_2 = \E -
(1-m)\K,\quad b_3 = - m\E,\quad b_4 = 2(r_2-r_1)(1-m)\K 
\eea
\ese

Importantly, 
the modulation system \eqref{e:KPWS} contains six PDEs for the five dependent variables $r_1$, $r_2$, $r_3$, $q$ and~$p$,
and is therefore overdetermined in general.
In~\cite{Biondini_2017}, the initial value problem for the system~\eqref{e:KPWS} was
shown to be compatible 
provided that \eqref{e:KPWS-3} and~\eqref{e:qconstraint} hold at $t = 0$, 
in which case it was shown that \eqref{e:KPWS-3} and~\eqref{e:qconstraint} remain satisfied 
for all $t > 0$.  
Consequently, 
in \cite{Biondini_2017}
a reduced system consisting of the five PDEs~\eqref{e:KPWS-1}--\eqref{e:KPWS-3}
was considered, which is 
a minimal set of equations for the five dependent variables 
$\@r = (r_1,r_2,r_3,q,p)^T$ that can be written as
\be
I_4\,\partialderiv{\@r}{t} + A_5\,\partialderiv{\@r}{x} + B_5\,\partialderiv{\@r}{y} = 0\,,
\label{e:KPWS5}
\ee
where $I_4 = \diag(1,1,1,1,0)$ and $A_5$ and $B_5$ are $5\times 5$
matrices whose explicit form is given in \eqref{e:A5x5&B5x5}.
In \cite{Biondini_2017} and \cite{Biondini_2020} the term ``KP-Whitham system''
was used to refer to the five equations~\eqref{e:KPWS5}.
However, in this work we will show that,
in order for the modulation system to inherit the integrability properties of the KP equation,
it is crucial to consider all six equations~\eqref{e:KPWS} 
on an equal footing.
Accordingly, we will henceforth refer to the system~\eqref{e:KPWS5} 
as the ``original KPWS'',
and we will refer to the six equations~\eqref{e:KPWS} as the ``full KPWS''.

Generally, all the dependent variables in~\eqref{e:KPWS} depend on two spatial dimensions ($x$ and $y$) 
and one temporal dimension ($t$), 
so we refer to the KPWS~\eqref{e:KPWS} as $(2+1)$-dimensional, or equivalently 3-dimensional.  
A number of asymptotic reductions of the
system~\eqref{e:KPWS} and their properties were studied
in~\cite{Biondini_2020}, and $(1+1)$-dimensional (i.e., 2-dimensional) reductions of the
soliton limit of~\eqref{e:KPWS} were used in \cite{NLTY21,PRSA22,JFM21} to
study various concrete physical problems.  
A number of important questions remain open, however.  
Among them is the issue of integrability.  
Since the KP-Whitham system was derived as an exact asymptotic reduction of the KP equation~\eqref{e:KPo}, which is integrable, 
one would naturally expect that the modulation system is also integrable.  
On the other hand, as was mentioned in
\cite{Biondini_2017}, the system~(\ref{e:KPWS}a--c) fails the Haantjes
tensor test for integrability \cite{MathAnn2007}.  
At the same time,
it was shown in \cite{Biondini_2020} that the harmonic and soliton
limits of the system~\eqref{e:KPWS} are in fact integrable.  An
obvious question is then whether there are other integrable reductions
of~\eqref{e:KPWS} and if so how one can identify them.

In this work, we begin to address this question by studying and
characterizing the two-dimensional ($1+1$ and $2+0$) reductions of the
KPWS~\eqref{e:KPWS}.  We demonstrate that these reductions of the
original KPWS are integrable only if the full KPWS is compatible.  In
section~\ref{s:yt} we consider the situation in which all fields are
independent of $x$, and in section~\ref{s:xt} the situation in which
all fields are independent of~$y$.  In section~\ref{s:xy} we study the
situation in which all fields are stationary, i.e., independent of~$t$.  
Finally, in section~\ref{s5remarks} we discuss the
compatbility of the full KPWS, and in section~\ref{s:conclusions} we
conclude this work with some final remarks.  
We emphasize that, as
in \cite{NLTY21,JFM21,PRSA22}, even when the solution of the KPWS is
independent of one independent variable, the reduced systems of
equations still generically describe two-dimensional, dynamical
configurations of the KP equation, because nonzero $V$ \eqref{e:qV}
implies propagation of the cnoidal wave and $q$ describes the
orientation of the periodic wave in the $xy$ plane.  Variations of $q$
with respect to $x$ or $y$ correspond to curved wave profiles.
Section~\ref{s:conclusions} ends this work with some concluding
remarks.

To avoid confusion, we note that in this work we are using the normalization of \cite{Biondini_2017}, not that of
\cite{Biondini_2020,PRSA22,NLTY21,JFM21}.  
In the latter works, the coefficient 6 in front of the term $uu_x$ in~\eqref{e:KP} was absent
and the cnoidal wave's period was normalized to $2\pi$, 
whereas here it is normalized to unity.  
As a result, several formulas are adjusted accordingly.

\section{The YT system} 
\label{s:yt}

In this section we consider solutions of the KPWS in which all fields are independent of $x$. 
We begin by considering the original KPWS~\eqref{e:KPWS5} 
[i.e., the five-component system~(\ref{e:KPWS}a--c)], neglecting the compatibility condition~\eqref{e:qconstraint} at first.
When solutions are independent of $y$, \eqref{e:KPWS5} reduces
to a system of five PDEs in the independent variables $y$ and $t$, 
which we refer to as the ``YT system''.  In vector form, this YT system is 
\be
\label{e:KPWS_yt}
I_4\partialderiv{\@r}{t} + B_5\partialderiv{\@r}{y} = 0\,,
\ee
where $\@r=(r_1,r_2,r_3,q,p)^T$ and $I_4 = \diag(1,1,1,1,0)$ as before,
and the coefficient matrix $B_5$ is given in~\eqref{e:B5def}.

\subsection{Reduction of the YT system to a three-component system}

The number of equations in the system \eqref{e:KPWS_yt} can be reduced through a suitable change of variables. 
Explicitly, the last row of \eqref{e:KPWS_yt} is
\be
\partialderiv{r_1}{y} +\alpha\partialderiv{}{y}(r_3-r_1) = 0 \, .
\label{e:KPWS_yt_s_constraint}
\ee
Then, we use the transformation
\be
\label{e:s_var_transform}
s_1 = r_3-r_2\,,\quad s_2 = r_3-r_1\,,\quad s_3 = r_2-r_1\,
\ee
where we made the choice not to define these variables in cyclic fashion in order to preserve the property that 
$s_j\ge0$ $\forall j=1,2,3$ when the Riemann-type variables $r_1,\dots,r_3$ are well-ordered. 
This transformation leads to the set of equations
\bse
\label{e:KPWS_yt_s}
\bea
\partialderiv{s_j}{t} + 2\sigma q\partialderiv{s_j}{y}+\sigma \Delta\nu_j \partialderiv{q}{y}=0 \,,
\quad 
j=1,2,3,
\label{e:KPWS_yt_sj}\\
\partialderiv{q}{t} +6\partialderiv{r_1}{y} + (\nu_4 + 2)\partialderiv{s_2}{y} +2\sigma q\partialderiv{q}{y} =0 \,,
\label{e:KPWS_yt_q}
\eea
\ese
with $\Delta\nu_1 = \nu_3 - \nu_2$, $\Delta\nu_2 = \nu_3 - \nu_1$ and $\Delta\nu_3 = \nu_2 -\nu_1$.
The transformation \eqref{e:s_var_transform} is not invertible. 
However, \eqref{e:KPWS_yt_sj} with $j=1$ 
is decoupled from the rest of the system, 
since $s_1$ does not appear in the remaining equations. 
Thus we can simply disregard it 
moving forward, 
since the four dependent variables $r_1$, $s_2$, $s_3$ and $q$,
determined by the PDEs \eqref{e:KPWS_yt_sj} with $j=2,3$ plus 
\eqref{e:KPWS_yt_q}, are a closed system.  These dependent variables, together with~\eqref{e:KPWS_yt_s_constraint}, are sufficient to recover the solution of the KP equation.

Next, one can use~\eqref{e:KPWS_yt_s_constraint} to eliminate $r_1$ from~\eqref{e:KPWS_yt_q},
obtaining the following closed system of three PDEs for the three dependent variables $s_2$, $s_3$ and~$q$:
\bse
\label{e:KPWS_yt_s2s3q}
\bea
\label{e:KPWS_yt_s_2}
\partialderiv{s_2}{t} + 2\sigma q\partialderiv{s_2}{y}+\sigma (\nu_3-\nu_1) \partialderiv{q}{y} = 0 \,, 
\label{e:KPWS_yt_s2_2}\\
\partialderiv{s_3}{t} + 2\sigma q\partialderiv{s_3}{y}+\sigma (\nu_2-\nu_1) \partialderiv{q}{y} = 0 \,,
\label{e:KPWS_yt_s3_2}\\
\partialderiv{q}{t} + (\nu_4  - 6\alpha + 2)\partialderiv{s_2}{y} +2\sigma q\partialderiv{q}{y} = 0 \,,
\label{e:KPWS_yt_q_2}
\eea
\ese
All the coefficients appearing in~\eqref{e:KPWS_yt_s2s3q} are completely determined by~$m$ and~$b$, 
which in turn are completely determined by $s_2,s_3\&q$ as
\be
\label{e:bmforms}
m = \frac{s_3}{s_2}\,,\qquad \frac bm = 2s_2\,.  \ee Note that $r_1$
is also needed to recover the asymptotic solution of the KP equation,
but its value, up to an integration constant determined by the initial
conditions, can be obtained from $s_2$, $s_3$ by
integrating~\eqref{e:KPWS_yt_s_constraint}.  Introducing
the vector $\@{v}= (s_2,s_3,q)^T$, we can write the above system
\eqref{e:KPWS_yt_s2s3q} in vector form as \be
\label{e:yt_3comp_reduction}
\partialderiv{\@{v}}{t} + B_3 \partialderiv{\@{v}}{y} = 0\,,
\ee
with 
\be
B_3 = \left( \begin{array}{ccc}
2\sigma q&0&\sigma(\nu_3-\nu_1)\\ 
0&2\sigma q&\sigma(\nu_2-\nu_1)\\
\nu_4-6\alpha+2&0&2\sigma q
\end{array} \right) 
\label{e:yt_3x3}\,.
\ee
The eigenvalues of $B_3$ are 
\bse
\label{e:YT_3comp_eigs}
\be 
\lambda_1 = 2 \sigma  q, \qquad \lambda_{2,3} = 2 \sigma  q \pm \sqrt{\Delta}\,,
\ee
where 
\be
\label{e:Delta}
\Delta = \sigma  \left(\nu _1-\nu_3\right) \left(6 \alpha -\nu _4-2\right)
  = 4\sigma s_2\frac{((1-m)\K -2 (2-m)\E\K + 3\E^2)^2}{3\E\K(\K-\E)(\E - (1-m)\K)}
  \,.
\ee
\ese
By properties of $K_m$ and $E_m$, $\sgn\Delta = \sgn\sigma$ since $s_2>0$.
Hence, if $\sigma=-1$, as for KPI, some of the eigenvalues are
  imaginary, implying that the initial value problem for the above system 
is ill-posed, confirming known results \cite{Biondini_2017}.
Incidentally, note that the PDEs for $s_2$ and $q$ do not contain~$s_3$ explicitly.
However, the value of $s_3$ is required to determine~$m$.

\subsection{Harmonic and soliton limits of the YT system} 
\label{s:XInd_Limits}

We now consider two distinguished limits of the system outlined in
\eqref{e:KPWS_yt_s_2}, which describe concrete physical scenarios: the
harmonic limit and the soliton limit.  The limiting values of all
coefficients of the KPWS in these limits are found in the Appendix.

The harmonic limit, 
in which the elliptic parameter $m \to 0$,
is obtained by taking $r_2\to r_1^+$. 
In this limit, the cnoidal wave becomes a vanishing-amplitude trigonometric wave
(cf.\ section~\ref{s:background} and the well-known properties of the elliptic functions \cite{NIST}).
Note $s_3\to0$ in this limit, consistent with the fact that $m$ is obtained 
from $s_2$ and $s_3$ via \eqref{e:bmforms}.
The harmonic limit of system \eqref{e:KPWS_yt_s_2} results in the partially decoupled system
\bse
\label{e:KPWS_yt_m0}
\bea
\partialderiv{s_2}{t} + 2\sigma q\partialderiv{s_2}{y} =0 \label{e:KPWS_yt_m0_s2}\,,\\
\partialderiv{q}{t}  +2\sigma q\partialderiv{q}{y} =0 \label{e:KPWS_yt_m0_q} \,,
\eea
\ese
with identical equations for the variables $s_2\& q$. 
In the limit $m \to 0$ of \eqref{e:KP_cn_wave},  
the mean flow is simply $\bar{u}=r_3$.
Moreover, \eqref{e:KPWS_yt_s_constraint} implies that, in the harmonic
limit, $\partial{r_3}/\partial{y} \to 0$.
Thus the mean flow is constant with respect to all spatial variables,
but it does not play any role in the system. 
Moreover, taking into account the reduction of~\eqref{e:KPWS_yt_s_constraint}, 
we see that the system \eqref{e:KPWS_yt_m0} coincides with the harmonic limits 
in \cite{Biondini_2020} 
and \cite{Biondini_2017} 
when derivatives with respect to $x$ are neglected.

The soliton limit, i.e., $m\to1$, is obtained when $r_2\to r_3^-$,
which amounts to $s_3\to s_2$, and
the cnoidal wave solution~\eqref{e:KP_cn_wave} limiting to a line soliton.
The first two equations of~\eqref{e:yt_3comp_reduction} are identical in this limit, consistent with $s_3 = s_2$,
while the remaining equations are
\bse
\label{e:KPWS_yt_m1}
\bea
\partialderiv{s_2}{t} + 2\sigma q\partialderiv{s_2}{y}+ \frac{4}{3}\sigma s_2\partialderiv{q}{y}=0 \label{e:KPWS_yt_m1_s2}\\
\partialderiv{q}{t} + 4\partialderiv{s_2}{y} +2\sigma q\partialderiv{q}{y} =0 \label{e:KPWS_yt_m1_q}\,.
\eea
\ese
As before, 
\eqref{e:KPWS_yt_m1} coincides with the reduction of the soliton limit in \cite{Biondini_2017} 
and \cite{Biondini_2020} 
when derivatives with respect to $x$ are neglected.
In this case, since $\alpha\to0$ as $m\to1$, \eqref{e:KPWS_yt_s_constraint} implies 
$\partial r_1/\partial y = 0$ and $\partial{s_2}/\partial{y}=\partial{r_3}/\partial{y}$.
Unlike the harmonic limit, the two-component
system~\eqref{e:KPWS_yt_m1} in the soliton limit is coupled. 
On the other hand, the system can be diagonalized in a straightforward way.
The eigenvalues of the system and the associated left eigenvectors
are, respectively \cite{Biondini_2020},
\be
\lambda_{\pm} = 2\sigma q \pm 4\sqrt{{\sigma s_2}/{3}}\,,
\qquad
\@{w}_\pm = ( \pm\sqrt{3}, \sqrt{\sigma s_2} )\,.
\ee
Using these, we obtain the characteristic differential forms 
$\sqrt{3}\d s_2 \pm \sqrt{\sigma s_2}\d q = \sqrt{3\sigma/s_2}\d s_2 \pm \d q = 0$,
which yields the Riemann invariants
\be
R_\pm = q\pm 2\sqrt{3\sigma s_2}\,.
\ee
In turn, the change of variable from $s_2$ and $q$ to $R_\pm$ transforms the system~\eqref{e:KPWS_yt_m1} 
into diagonal form:
\be
\partialderiv{R_\pm}{t} + \lambda_\pm \partialderiv{R_\pm}{x} = 0\,.
\ee

\subsection{Integrability and Riemann invariant of the YT system}
\label{e:yt_diag}

We now return to the three-component YT
system~\eqref{e:yt_3comp_reduction}.  The Haantjes tensor test for
integrability of a system of hydrodynamic equations \cite{MathAnn2007}
(see also the Appendix) is a relatively simple way to determine
whether a strictly hyperbolic system is diagonalizable.  It is
generally believed that asymptotic reductions of an integrable system
preserve integrability.  While the KP equation~\eqref{e:KP} is
integrable, the five-component original KP Whitham system
(\ref{e:KPWS}a--c) fails the Haantjies tensor test, a necessary
condition for the integrability of three-dimensional quasi-linear
systems \cite{Ferapontov2006}. At the same time, it was shown in
\cite{Biondini_2020} that the three-dimensional harmonic and soliton
limits of the full system are indeed integrable, which then raises the
natural question of whether two-dimensional reductions of the KPWS are
integrable.

The three-component YT system~\eqref{e:yt_3comp_reduction} fails the
Haantjes tensor test, since 19 out of the 27 components of the
Haantjes tensor vanish in general.  The Haantjes tensor does vanish in
the limit $m\to0$, but its entries diverge in the limit $m\to1$, even
though, as we showed above, the limiting system can be reduced to an
integrable two-component system for $s_2$ and~$q$.

The reason for this discrepancy, and the reason why the original KPWS
and the reduction \eqref{e:yt_3comp_reduction} fails the Haantjes
tensor test is that, in order for \eqref{e:KPWS} or \eqref{e:KPWS5} to be
compatible, $q$ and $k$ must be related by
condition~\eqref{e:waveconservation3}, i.e., $k_y = l_x$, or,
equivalently, (\ref{e:KPWS}d). When the initial conditions do not
satisfy the compatibility condition (\ref{e:KPWS}d), the original KPWS
does not describe actual solutions of the KP equation, and that
explains why the original KPWS system may not be integrable despite
the integrability of the KP equation.  Indeed, we show below that,
once the compatibility condition $k_y = l_x$ is enforced, the
$x$-independent reduction of the full KPWS (the YT system) is in fact
integrable.  Later, in sections~\ref{s:xt} and~\ref{s:xy}, we will see
how similar considerations apply to the $y$-independent and
$t$-independent reductions of the KPWS.

The coefficient matrix $A$ in~\eqref{e:B5def} has the eigenvalue
$2\sigma q$, which is inherited by the matrix $B$ in~\eqref{e:yt_3x3}.
Using the corresponding left eigenvector in~\eqref{e:YT_3comp_eigs},
we have the following characteristic relation
\be
(\nu_2-\nu_1)\d s_2-(\nu_3-\nu_1)\d s_3 = 0\,,
\ee
along $\d y/\d t = 2\sigma q$.
To integrate this differential form,
we eliminate~$s_3$ in favor of $m$ using~\eqref{e:bmforms},
to obtain
\be
(\nu_2-\nu_1-m(\nu_3-\nu_1))\,\d s_2-s_2(\nu_3-\nu_1)\,\d m = 0\,.
\ee
Multiplying by the integrating factor $1/[s_2(\nu_2-\nu_1-m(\nu_3-\nu_1))]$ yields
\be
\frac{1}{s_2}\d s_2+\bigg(\frac{1}{m}-\frac{E_m}{m(1-m)K_m}\bigg)\d m=0\,,
\ee
and recalling the derivative of $\K$ [cf.~\eqref{e:dkdm_ODE}], 
we express the above characteristic relation as
\be
\d\big[ \frac{1}{2} \log s_2 - \log \K \big] = 0,
\ee
which yields the Riemann invariant
$R_o = {\sqrt{s_2}}\,\big/{2\K}$
that satisfies the PDE
\be
\partialderiv{R_o}{t}+2\sigma q\partialderiv{R_o}{y}=0\,.
\ee
In light of~\eqref{e:def_k} and~\eqref{e:s_var_transform}, however, we see that $R_o\equiv k$!  
That is, the Riemann invariant $R_o$ is just the local wavenumber $k$ in the $x$ direction.
Moreover, the fact that $k$ is a Riemann invariant is deeply connected
with the integrability of the full KPWS.  This is because, when all
fields are independent of~$x$, the conservation of waves and
compatibility condition~\eqref{e:waveconservation} immediately yield
$k_t = k_y = 0$, i.e., $k={}$const.  Enforcing the constancy of the
local wavenumber $k$ (as needed to ensure the compatibility of the
full KPWS with the KP equation, as per the above discussion) reduces
the three-component YT system~\eqref{e:yt_3comp_reduction} to a
two-component system that is locally integrable.  Any two-component
system can always be reduced to Riemann invariant form and is
therefore always locally integrable via the classical hodograph
transform.  We will see in sections~\ref{s:xt} and~\ref{s:xy} that a
similar phenomenon will also arise for the $y$-independent and
$t$-independent reductions of the KPWS.

\subsection{Further reduction of the YT system and its diagonalization}
\label{s:s2_q_syst}

Since the wavenumber $k$ is simultaneously a Riemann invariant of the
three-component YT system~\eqref{e:yt_3comp_reduction} and a conserved
quantity for $x$-independent one-phase solutions of the KP equation,
we now derive a reduction of the YT system in which $k$ is
constant.  We choose two different sets of dependent variables: a
reduced system for the 
dependent variables $(s_2,q)^T$ and the reduced system for the
dependent variables $(m,q)^T$.

We begin by performing a change of variable from $\@{v}=(s_2,s_3,q)^T$
to $\~{\@v}=(s_2,q,k)^T$.  We choose to keep $s_2$ as opposed to $s_3$
because $s_2$ never vanishes, whereas $s_3\to0$ in the harmonic
limit.  Then $\~{\@v}$ satisfies the system $\~{\@v}_t +
\~B_3\,\~{\@v}_y = 0$, with $\~B_3 = T^{-1}B_3T$ and $T = \big(
{\partial v_i}/{\partial\~v_j} \big)$\,.  One can
verify that the new coefficient matrix is block-diagonal,
$\~B_3 = \diag(B_2,2\sigma q)$, with the $2\times2$ matrix $B_2$ given
by
\bea
B_2 = \left(\! \begin{array}{cc} 
  2\sigma q & 2\sigma s_2 \displaystyle\frac{(1-m) \K^2 - 2 (2-m) \K\E + 3 \E^2}{3\E (\E-\K)} \\
  \displaystyle 2 + 2\left(\frac{m}{\E-(1-m)\K}-\frac{3}{\K}\right)\E  & 2\sigma q
  \end{array} \!\right)\,.
\nonumber\\
\label{e:Qyt2x2}
\eea Since compatibility of the KPWS requires $k_t=k_y=0$, we
therefore consider the compatible solution $k\equiv{}$const, which
simplifies the $3\times 3$ system $\~{\@v}_t + \~B_3\,\~{\@v}_y = 0$
to the following $2\times 2$ system for the dependent variable $\@u
= (s_2,q)^T$: \be \@ u_t + B_2 \,\@ u_x = 0\,.
\label{e:u2x2yt}
\ee
The coefficient matrix $B_2$ in~\eqref{e:Qyt2x2} contains the elliptic parameter $m$, 
which is defined implicitly in terms of $s_2$ and $k$ by the relation~\eqref{e:def_k}, implying
\be
\sqrt{s_2} = 2 k_0 \K\,,
\label{e:k_via_s2}
\ee
where $k_0$ is the positive constant value of $k$. We can solve for $m$ by inverting $\K$ via $m =
\K^{-1}(\sqrt{s_2}/(2k_0 ),m) = 1 - \mathrm{dn}^2(\sqrt{s_2}/(2k_0 ),m)$,
where $\mathrm{dn}$ is a Jacobi elliptic function.

In light of~\eqref{e:k_via_s2}, we see that \eqref{e:u2x2yt} is
actually a one-parameter family of hydrodynamic type systems,
parametrized by the constant value of $k_0 $ via~\eqref{e:k_via_s2}. 
Equivalently, one can use~\eqref{e:k_via_s2} to express $s_2$ as a function of $m$ and~$k_0 $.
Note however that $q$ does not enter in the relation between $
s_2$ and~$m$.  
One can therefore easily replace~\eqref{e:u2x2yt} 
with the equivalent hydrodynamic system $\~{\@u}_t +\~B_2 \,\~{\@u}_x = 0$ for the 
modified dependent variable $\~{\@u} = (m,q)^T$.

\begin{figure}[t!]
	\centerline{%
		\includegraphics[width=0.325\textwidth]{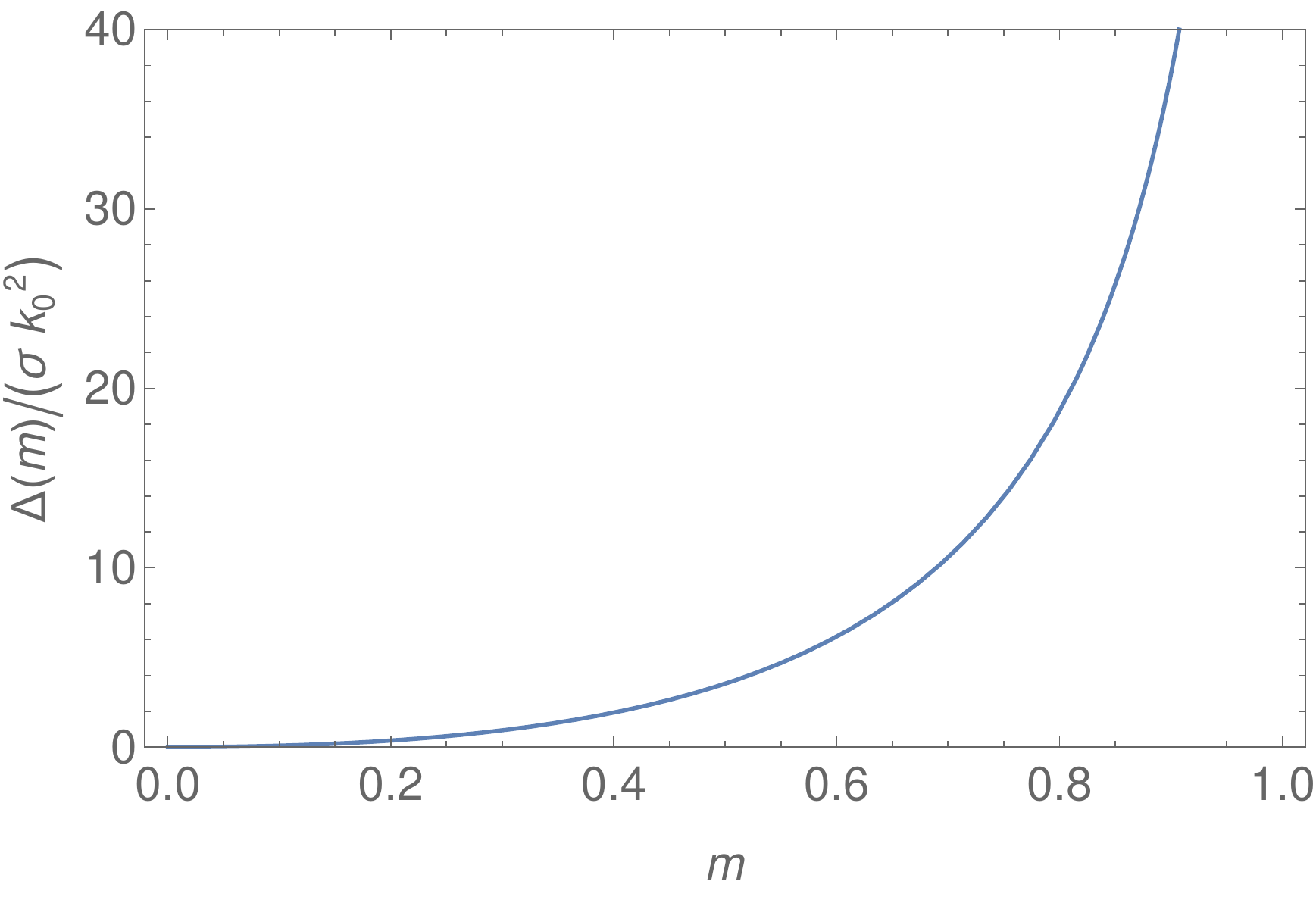}~
		\includegraphics[width=0.325\textwidth]{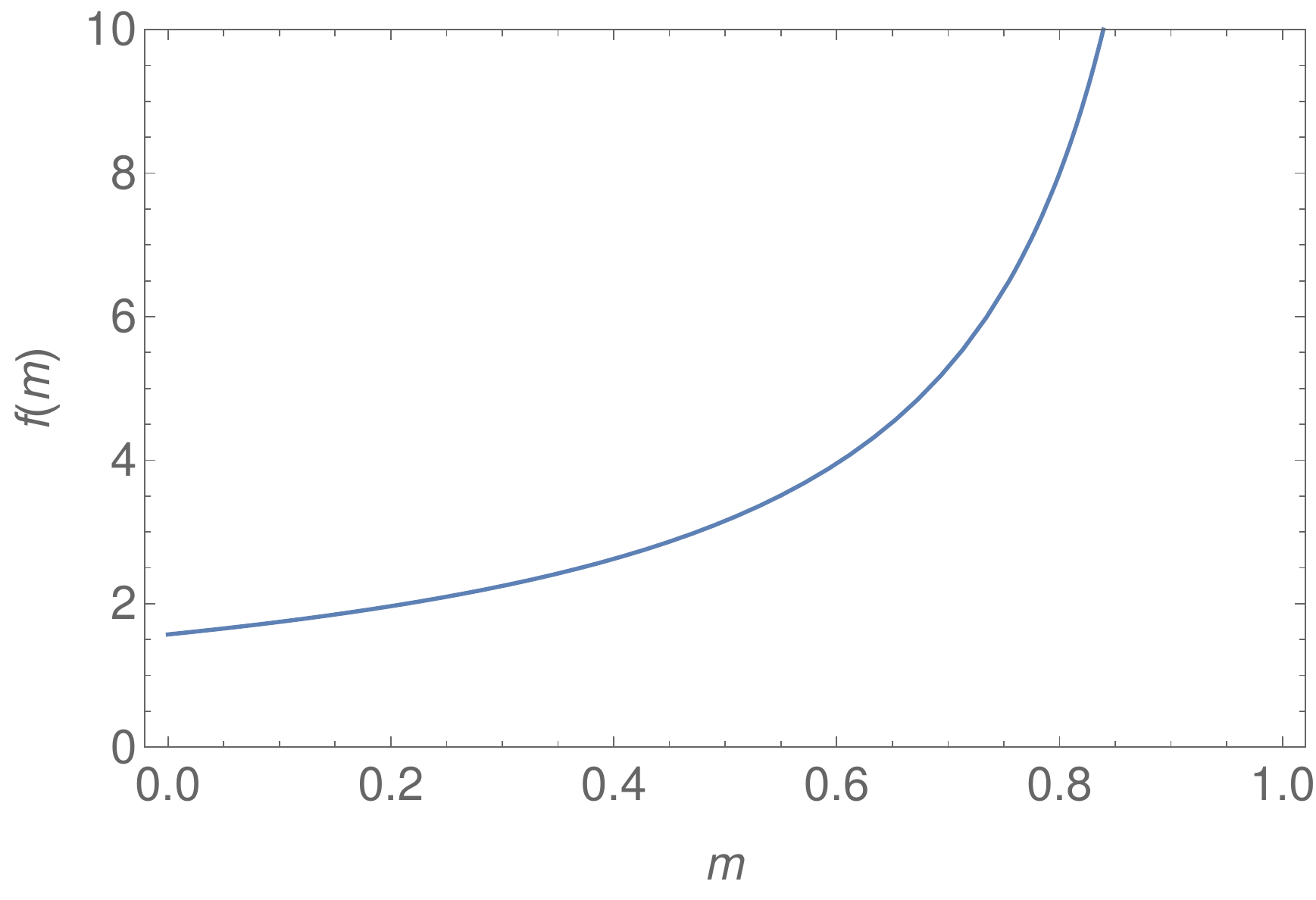}~~~
		\includegraphics[width=0.32\textwidth]{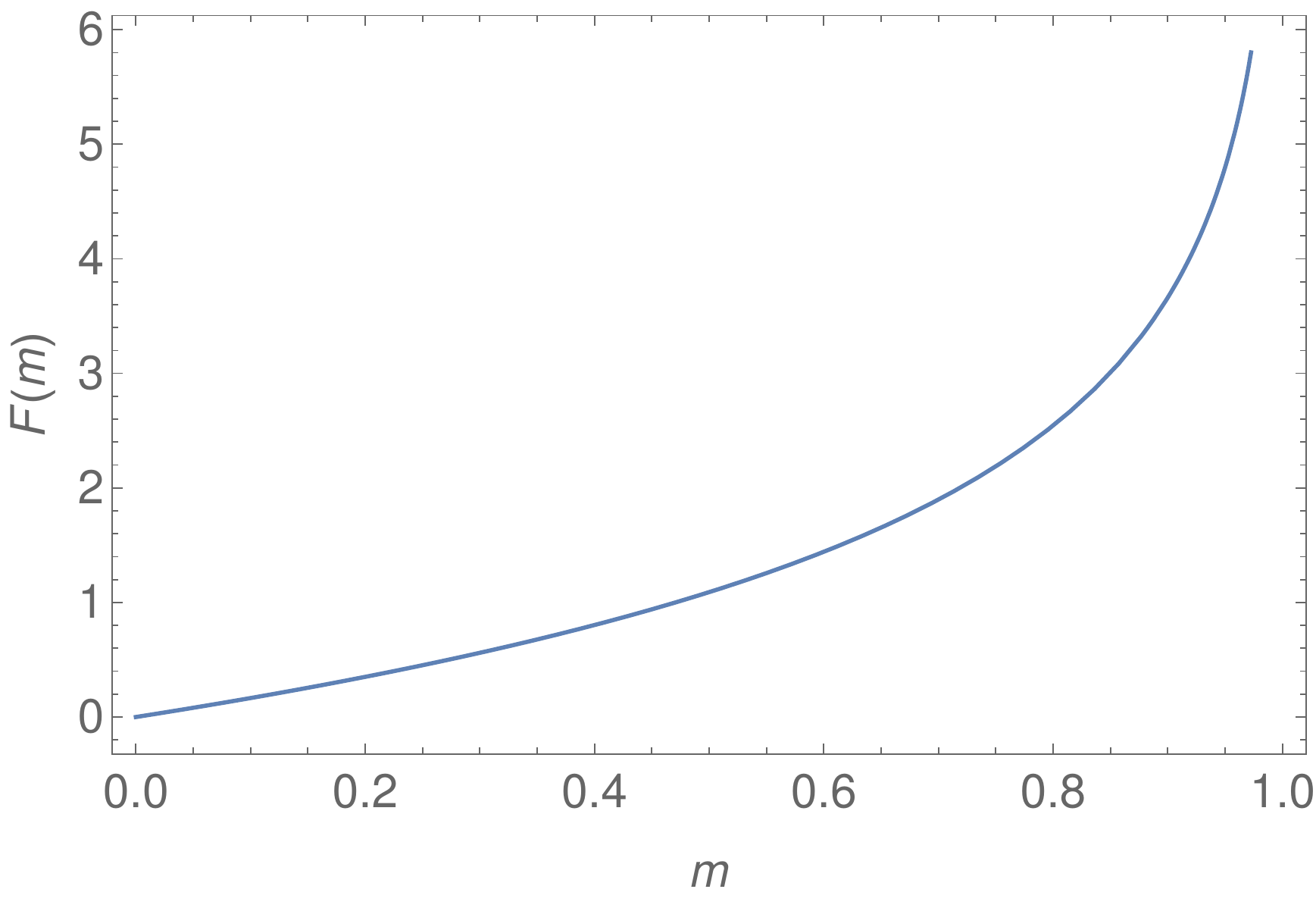}%
		}
	\caption{%
		The quantities $\Delta(m)/(\sigma k_0 ^2)$ (left), $f(m)$ (center) and $F(m)$ (right) as functions of~$m$. 
		}
	\label{f:YT_f&F}
\end{figure}

We use the eigenvalues and eigenvectors of~$B_2$ to complete the diagonalization of the YT system. 
The eigenvalues $\lambda_\pm$ coincide with $\lambda_{2,3}$ in~\eqref{e:YT_3comp_eigs},
while the associated left eigenvectors are
\be
\@w_\pm = \big( \pm \sqrt{3\sigma\E(\K-\E)/[s_2\K(\E-(1-m)\K))]} \,,\, 1 \,\big)\,,
\ee
leading to the characteristic relations
\bse
\be
\d q \pm \sqrt{\frac{3\sigma\E(\K-\E)}{s_2\K(\E-(1-m)\K)}} \d s_2  = 0
\label{e:YT2x2_characteristic_relation}
\ee
along the characteristic curves 
\be
\d y/\d t = 2\sigma q \mp \sqrt{\Delta}\,,
\label{e:ytcharacteristiccurves}
\ee
\ese
with 
$\Delta$ as in~\eqref{e:YT_3comp_eigs}.
We now differentiate~\eqref{e:k_via_s2} and use the known differential equations for $\K$, \eqref{e:dkdm_ODE},
to express 
$\d s_2 = s_2[(\E - (1-m)\K)/(m(1-m)\K)]\,\d m$, 
thereby simplifying~\eqref{e:YT2x2_characteristic_relation} to
\bse
\be
\d q \pm \sqrt\sigma k_0  f(m)\,\d m = 0\,,
\ee
where
\be
f(m) = \frac{2\sqrt{3\E(\K-\E)(\E - (1-m)\K)}}{m(1-m)\sqrt{\K}}\,.
\ee
\ese
Therefore, Riemann invariants for the two-component hydrodynamic system for $\~{\@u}$ are 
\bse
\be
R_\pm = q \pm \sqrt\sigma k_0 \,F(m)\,,
\label{e:yt_2x2Riemanninvariants}
\ee
with
\be
F(m) = \int_0^m f(\mu)\,\d\mu\,.
\ee
\ese
Although we were unable to compute this integral in closed form,
the above expression of the Riemann invariants is the same as those
for the p-system modeling isentropic gas dynamics and nonlinear
elasticity \cite{Smoller1994} where $f(m)$ is related to the sound
speed of the medium.
A plot of $f(m)$ is shown in Fig.~\ref{f:YT_f&F}.
Note that $f(m)>0$ for all $m\in[0,1)$, and 
$\lim_{m\to0^+}f(m) = \pi/2$. 
However, $f(m) = 2[1 + 2/(\log(1-m)-4\log2)]^{1/2}/(1-m) + O(1)$ as $m\to1^-$.
As a result, we also have $F(m) \to+\infty$ and $R_\pm\to\pm\infty$
logarithmically in that limit, 
implying that the limit $m\to1^-$ of the two-component system for $m$
and~$q$ is singular. 
This is expected because $k_0  \equiv{}$ const $\ne0$ is incompatible with the 
soliton limit, for which $k\to0$ [cf.~\eqref{e:klomegadef}].
Note that $k_0 = 0$, $m \to 1$ is a valid reduction that corresponds
to the zero amplitude limit $s_2 \to 0$ for the soliton modulation
equations \eqref{e:KPWS_yt_m1_q}. In
Fig.~\ref{f:YT_f&F} we also plot 
$F(m)$ as a function of $m$, which can be used to obtain the relationship between $m$ and $q$ satisfied by
simple wave solutions of \eqref{e:u2x2yt} with $R_\pm \equiv {}$const.

In closing, 
we return our attention to the eigenvalues $\lambda_\pm$ of the system, given by $\lambda_\pm = 2\sigma q \pm \sqrt{\Delta(m)}$ [cf.~\eqref{e:YT_3comp_eigs}],
where in light of~\eqref{e:k_via_s2}, we can now express $\Delta$ as
\be
\Delta(m)
  = 16\sigma k_0 ^2\K\frac{((1-m)\K -2 (2-m)\E\K + 3\E^2)^2}{3\E(\K-\E)(\E - (1-m)\K)}\,.
\ee
Note that $\Delta(m)/(\sigma k_0 ^2)\to0$ as $m\to0$ and $\Delta(m)/(\sigma k_0 ^2)\to+\infty$ as $m\to1$
(cf.\ Fig.~\ref{f:YT_f&F}).  When $m = 0$,
$\lambda_\pm = 2 \sigma q$ 
and $R_\pm = q$ so that \eqref{e:u2x2yt} exhibits a one-component
reduction in the harmonic limit to the inviscid Burgers equation.  
By
monotonicity of $\Delta(m)$, the reduced $2\times2$
system is strictly hyperbolic according to the following definition of
strict hyperbolicity:  $\lambda_+ = \lambda_-$ if and only if $R_+ =
R_-$ \cite{ElHoefer}.
It can also be shown that
$(\grad_{m,q} \lambda_\pm) \cdot \@ w_\pm \neq 0$,
which implies that the system is genuinely nonlinear.

\section{The XT system}
\label{s:xt} 

Next we consider the reduction of the KPWS in which all fields are independent of~$y$. 
Similarly to section~\ref{s:yt}, at first we consider the
five-component original KPWS~\eqref{e:KPWS5},
ignoring the compatibility condition~\eqref{e:qconstraint}.
When solutions are independent of $y$, this 
system reduces to what we call the ``XT system'', 
which in vector form is 
\be
\label{e:KPWS_xt}
I_4\partialderiv{\@r}{t} + A_5\partialderiv{\@r}{x} = 0\,,
\ee
with $\@r = (r_1,r_2,r_3,q,p)^T$ as before, and 
$A_5$ given in~\eqref{e:A5def}.
We will obtain analogous results to section~\ref{s:yt}
even though the analysis of the systems and the physics they describe are significantly different.

\subsection{Reduction to a four-component system, harmonic and soliton limits}

In this case, reducing the size of the system is much easier than in section~\ref{s:yt}, because
when all derivatives in $y$ vanish,
the last equation in \eqref{e:KPWS_xt} determines $p$ in terms of
$r_1,r_2,r_3$ by direct integration.  
Substituting the resulting expression into the PDEs for $r_1,r_2,r_3$ leads to the reduced system
\bse
\label{e:KPWS_xt_2}
\bea
\pdv{r_j}{t}+\qty(V_j-\sigma q^2)\pdv{r_j}{x}
  + \sigma q^2(1-\alpha)\pdv{r_1}{x} + \sigma q^2\alpha\pdv{r_3}{x} 
  + \sigma q(\nu_5 -\nu_j)\pdv{q}{x} =0\,,\quad j=1,2,3\,,
\nonumber\\[-1ex]
\\
\pdv{q}{t} -q (4-\nu_4)\pdv{r_1}{x} + q(2+\nu_4)\pdv{r_3}{x} +(V_2-\sigma q^2)\pdv{q}{x} =0\,,
\eea
\ese
or equivalently, in vector form,
\be
\partialderiv{\@r_4}{t} + A_4 \partialderiv{\@r_4}{x} = 0\,, 
\label{e:xt_Vect}
\ee
where now $\@r_4 = (r_1,r_2,r_3,q)^T$ 
and
\be
A_4 = \left(\begin{array}{cccc}
V_1 - \sigma q^2 \alpha & 0&\sigma q^2\alpha&\sigma q (\nu_5-\nu_1)\\
\sigma q^2 (1-\alpha)& V_2 -\sigma q^2 &\sigma q^2 \alpha &\sigma q (\nu_5-\nu_2)\\
\sigma q^2 (1-\alpha) & 0& V_3 + \sigma q^2 (\alpha-1) & \sigma q (\nu_5-\nu_3)\\
-q(4-\nu_4) & 0 & -q(2+\nu_4) & V_2 -\sigma q^2
\end{array}\right)\,.
\ee

The above system simplifies considerably in the harmonic and soliton limits.
The limiting values of all coefficients are found in the Appendix. 
In the harmonic limit, $m\to0$, the PDEs for $r_1$ and $r_2$ coincide, and   
we can therefore choose the reduced set of dependent variables $\@r_3 = (r_1,r_3,q)^T$, obtaining
\be
\pdv{\@r_3}{t} + A_3^{(0)} \pdv{\@r_3}{x} = 0\,,
\label{e:xt_3comp_m=0&m=1}
\ee
with
\be
A_{3}^{(0)} = \left(\begin{array}{ccc}
12r_1-6r_3 - \sigma q^2  &\sigma q^2 &0\\
0 & 6r_3  & 0\\
0 & 6 & 12r_1-6r_3 -\sigma q^2
\end{array}\right)\,.
\ee
Equation~\eqref{e:xt_3comp_m=0&m=1} coincides with 
the $y$-independent reduction of the systems 
in \cite{Biondini_2020},
which were shown to be integrable. 
Conversely, in the soliton limit, $m\to1$, the PDEs for $r_2$ and $r_3$ coincide. 
Choosing again $\@r_3 = (r_1,r_3,q)^T$, we obtain~\eqref{e:xt_3comp_m=0&m=1},
but with $A_3^{(0)}$ replaced by
\be
A_{3}^{(1)} = \left(\begin{array}{ccc}
6r_1&0&0\\
\sigma q^2 & 2r_1+4r_3 - \sigma q^2 & \frac{4\sigma q}{3}(r_1-r_3)\\
-2q & -4q & 2r_1+4r_3 -\sigma q^2
\end{array}\right)\,.
\ee
This system 
also coincides with the $y$-independent reduction of the systems 
in \cite{Biondini_2020}, 
and is therefore integrable.

\subsection{Riemann invariant, integrability, further reduction and diagonalization of the XT system}
\label{s:xt_diagonalization}

The matrix $A_4$ has the eigenvalue
\be
\lambda_o = V_2 -\sigma q^2\,,
\label{e:xt_simple_eig}
\ee
with associated left eigenvector
\bea
\@w_o = \big( q(1-m)(\K-\E)\,,\, q(\E-(1-m)\K)\,,\, qm\E\,, \, - 2m(1-m)(r_3-r_1)\K \,\big)\,.
\nonumber\\[-1ex]
\label{e:xt_simple_vect}
\eea
These expressions allow us to find a Riemann invariant and, in turn, to 
partially diagonalize the XT system~\eqref{e:xt_Vect}. 
In this case however, the calculations are more complicated than those
of section~\ref{s:yt}.
We begin by applying $\@w_o^T$ to~\eqref{e:xt_Vect}, obtaining the characteristic relation
\be
\frac1{(r_3-r_1)\K} \bigg( \frac{\K-\E}{m}\d r_1
  + \frac{\E - (1-m)\K}{(1-m) m}\d r_2
  - \frac{\E}{1-m} \d r_3 \bigg) 
  - 2 \frac{\d q}q = 0\,
\ee
along $\d{x}/\d{t}=V_2 -\sigma q^2$\,.
To integrate this differential form and find the Riemann invariant, 
we first eliminate $r_2$ in favor of $m$ using \eqref{e:m_def}, 
implying
$\d r_2 =(1-m)\,\d r_1 +m\,\d r_3 + (r_3-r_1)\,\d m$,
which yields
\be
\frac{\d(r_3-r_1)}{r_3-r_1} + \frac{\E - (1-m)\K}{m(1-m)\K}\, \d m - 2\frac{\d q}q = 0\,.
\ee
Next, we eliminate $q$ in favor of $l$ using \eqref{e:qdef}, 
implying 
$\d q = \d l/k - l\,\d k /k^2$, 
and we use \eqref{e:def_k} to replace $k$ and express $\d k = \d(r_3-r_1)/(4\K\sqrt{r_3-r_1}) - \sqrt{r_3-r_1}\d\K/(2\K^2)$.
Using equation \eqref{e:dkdm_ODE},
substituting and simplifying, the resulting expression finally yields
$\d l = 0$.  The Riemann invariant $R_o$ in this case is nothing else but
the wavenumber in the $y$ direction,
$R_o = l$,
which satisfies the PDE
\be
\partialderiv{R_o}t + (V_2 - \sigma q^2)\partialderiv{R_o}x = 0\,.
\ee

The fact that $l$ is a Riemann invariant for the XT system is not an accident, 
and is related to its compatibility with the full KPWS and (as we will see below) with its integrability.
This is because, when all fields are independent of $y$, the conservation of waves equations~\eqref{e:waveconservation}
yield $l_x = l_t = 0$, implying that, for one-phase solutions of the KP equation, $l$ must be constant. 
We can use this relation to reduce the XT system~\eqref{e:xt_Vect}
to a three-component system.
To do this, we perform a change of dependent variable from $\@r_4 = (r_1,r_2,r_3,q)^T$ to $\@v = (r_1,r_2,r_3,l)^T$,
which results in the partially decoupled system
\be
\@v_t + A_4'\,\@v_x = 0\,,
\label{e:xt_4x4partialdecouple}
\ee
with 
\be
A_4' = \left(\begin{array}{cc} A_3 & \@ a_3 \\ \@0_3^T & V_2 - \sigma q^2 \end{array}\right)\,,
\qquad
\ee
where the three-component vector $\@a_3$ is immaterial for our purposes, $\@0_3 = (0,0,0)^T$ and
\unskip%
\bse
\label{e:xt_A3x3}
\bea
A_3 = A_\mathrm{KdV} + \frac{4\sigma\K l^2}{3(r_2-r_1)} A_3^{(2)}\,,
\\
A_\mathrm{KdV} = \diag(V_1,V_2,V_3)\,,
\\
A_3^{(2)} = \left(\begin{array}{ccc}
  2a_3
    & \frac{a_1a_4}{(1-m)(\K-\E)} 
    & \frac{m \E a_2}{(1-m)(\K-\E)} \\[0.4ex]
 \frac{-(\K-\E)a_4}{a_1} 
    & -\frac{2a_2}{1-m} 
    & - \frac{\E m a_3}{(1-m)a_1} \\[0.4ex]
 \frac{(\K-\E)a_2}{\E} 
    & \frac{a_1a_3}{(1-m)\E} 
    & -\frac{2 m a_4}{1-m}
   \end{array}\right)\,,
\eea
with $V_1,\dots,V_3$ as in~\eqref{e:Vdef} and 
\bea
a_1 = \E - (1-m)\K\,,\quad
a_2 = (1-m)\K+(2m-1)\E\,,\\
a_3 = (1-m)\K - (1+m)\E\,,\quad
a_4 = 2 (1-m) \K - (2-m) \E \,.
\eea
\ese

Since $l$ is constant for compatible solutions of the full KPWS,
we can solve the fourth equation in~\eqref{e:xt_4x4partialdecouple} by
taking $l \equiv l_0$, 
thereby arriving at a three-component system of the same form
as~\eqref{e:xt_3comp_m=0&m=1}
\be
\pdv{\@r_3}{t} + A_3\,\pdv{\@r_3}{x} = 0\,,
\label{e:xt_3comp}
\ee
except that now 
$\@r_3 = (r_1,r_2,r_3)^T$ and the coefficient matrix $A_3$ is given by~\eqref{e:xt_A3x3}.
The case 
$l_0=0$ yields $A_3 = A_\mathrm{KdV}$, so
the system~\eqref{e:xt_3comp} reduces exactly to the 
KdV-Whitham system of Whitham modulation equations for the KdV equation \cite{Whitham1965}.
We have therefore showed that, once compatibility is enforced,
the XT~system reduces to a one-parameter deformation of the KdV-Whitham system, 
parametrized by the value of the wavenumber~$l_0$ along the transverse dimension. 

We now turn to the issue of the integrability of the XT~system.  
We apply the Haantjes tensor test to 
the four-component XT~system~\eqref{e:KPWS_xt_2} and find that it fails.
At the same time, the harmonic and soliton limits of the XT~system discussed earlier are clearly integrable, since they are 
exact one-dimensional reductions of the the harmonic and soliton limits of the full KPWS, which were shown to be integrable
in \cite{Biondini_2020}.
On the other hand, the three-component reduced system
\eqref{e:xt_3comp} does pass the Haantjes test,
in that all the terms of its Haantjes tensor associated with $A_3$ in~\eqref{e:xt_A3x3} vanish identically.
Thus, while the system~\eqref{e:KPWS_xt} is not integrable,
the system~\eqref{e:xt_3comp} is an integrable, one-parameter family
of deformations of the KdV-Whitham system.

The harmonic limit (i.e., $r_2\to r_1$, implying $m\to0$) of the deformed three-component system~\eqref{e:xt_3comp} yields the coefficient matrix
\begin{equation}
\lim_{r_2 \to r_1} A_3 =
\left ( \begin{array}{ccc}
12r_1 - 6 r_3 - \sigma q^2 & 0 & \sigma q^2 \\ 0 & 12 r_1 - 6 r_3 -
                                                   \sigma q^2 & \sigma
                                                                q^2 \\
          0 & 0 & 6r_3
        \end{array} \right )
      \label{e:KdVdeformed_harmonic}
\end{equation}
with $q^2 = \pi^2 l_0^2/(r_3 - r_1)$, 
which is consistent with the system~\eqref{e:xt_3comp_m=0&m=1}.
On the other hand, like with the reduced YT system, the soliton limit (i.e., $r_2\to r_3$, implying $m\to1$) of the reduced 
XT system~\eqref{e:xt_3comp} is singular for $l_0 \ne 0$,
since some of the entries of $A_3^{(2)}$ diverge in that limit.
Again, this is to be expected, because in the soliton limit one has $l_0=0$ [cf.~\eqref{e:klomegadef}].

\subsection{Deformed Riemann invariants}
\label{s:xt_deformedinvariants}

Since the one-parameter deformation~\eqref{e:xt_3comp} of the KdV-Whitham system is integrable,
it can be written in diagonal form.
We have not been able to determine the deformed Riemann invariants for
all parameter values $l$ but we can obtain approximate Riemann
invariants for small $l$
following standard methodology (e.g., see \cite{hinch1991}). 
It is convenient to define the coefficient
\eq{
c_2 = \frac{4\K}{3m(r_2-r_1)}\,,
}
so that~\eqref{e:xt_A3x3} reads $A_3 = A_\mathrm{KdV} + A_\mathrm{def}$
and the ``deformation matrix'' is simply 
$A_\mathrm{def} = \ell c_2 A_3^{(2)}$ with the (signed) deformation
parameter
\bea
\ell \equiv \sigma l_0^2 .
\eea

We seek an expansion of the deformed eigenvalues $\~V_j$ and 
corresponding right eigenvectors $\~{\@v}_j$ 
in powers of $\ell$ as
\vspace*{0ex}
\bea
\~V_j &= V_j + \ell V_j ^{(2)}(\@r_3) + O(\ell^2),\qquad
\~{\@v}_j &= \@e_j  + \ell \@v_j ^{(2)}(\@r_3) + O(\ell^2),\qquad
j = 1,2,3,
\eea
where $\@r_3 = (r_1,r_2,r_3)^T$ are the unperturbed KdV-Whitham
Riemann invariants, 
the unperturbed speeds $V_j$ are given in \eqref{e:Vdef}, 
and the unperturbed eigenvectors $\@e_1,\@e_2,\@e_3$ are simply the canonical basis in $\Real^3$,
i.e., $(\@e_1,\@e_2,\@e_3) = \I_3$, where
$\I_n$ is the $n\times n$ identity matrix.
We begin by computing the perturbation to the characteristic speeds.  
Since $A_\mathrm{def} = O(\ell)$, the first correction terms appear at $O(\ell)$.
The deformed eigenvalue problem is
\vspace*{0ex}
\eq{
(A_\mathrm{KdV}+A_\mathrm{def}) \,\~{\@v}_j = \~V_j \~{\@v}_j\,,
\qquad j = 1,2,3.
\label{e:KdVdeformed_eval}
}
The unperturbed eigenvalue problem, obtained at $O(1)$, is simply
$A_\mathrm{KdV} \@e_i  = V_i \@e_i$, which is satisfied because the
KdV-Whitham system is the $\ell = 0 = l_0$ reduction of the XT system \eqref{e:xt_3comp}.
Collecting terms $O(\ell)$ in~\eqref{e:KdVdeformed_eval} yields
\eq{
\qty(A_\mathrm{KdV}-V_j\I_3)\@v_j ^{(2)} = \qty(V_j ^{(2)} - c_2 A_3^{(2)})\@e_j\,,
\label{e:firstorder}
}
and multiplying from the left by $\@e_j^T$ yields
the first-order correction to the characteristic velocities as
\eq{
V_j^{(2)} = \@e_j^T c_2 A_3^{(2)} \@e_j\,,\qquad
j=1,2,3\,,
\label{e:KdVdeformed_evalcorr}
}
since the $\@e_j$ are orthonormal.
Explicitly,
\bse
\label{e:KdVdeformed_evalapprox}
\bea
\~V_1 = V_1 ^{(0)}-2c_2 \ell\left((1+m)\E -(1-m)\K \right) + O(\ell^2),\nsub\\
\~V_2 = V_2 ^{(0)}-2c_2 \ell\frac{(1-m)\K-(1-2m)\E }{1-m} + O(\ell^2),\sub\\
\~V_3 = V_3 ^{(0)}+2c_2 \ell m\frac{(2-m)\E - 2(1-m)\K}{1-m} + O(\ell^2)\,.\sub
\eea
\ese
Note that if $\sigma = 1$, the corrections to the first and second characteristic
velocities are negative, 
while the correction to the third characteristic velocity is positive
(opposite if $\sigma = -1$).
Note also that all three corrections remain finite in the limit $m\to0$, but diverge as $m\to1$,
indicating that the harmonic limit of the deformed system is finite, 
but its soliton limit is singular for $\ell \ne 0$.  This is expected because the
soliton limit requires $l_0 = 0$, which is only true when the XT
system \eqref{e:xt_3comp} reduces to the KdV-Whitham system.
Next, we use~\eqref{e:firstorder} and~\eqref{e:KdVdeformed_evalcorr} to compute the correction to the eigenvectors.
The matrix $A_\mathrm{KdV}-V_j\I_3$ is singular for $j=1,2,3$.
Nonetheless, the inhomogeneous linear system admits a solution, which yields the deformed eigenvectors in the form
\bea
\vb{\~v_1} = \left(\!\!\begin{array}{c} 1\\ 0\\ 0 \end{array}\!\!\right) + c_2 \ell \left(\!\!\begin{array}{c} 0\\ 1\\ m \end{array}\!\!\right),\quad
\vb{\~v_2} = \left(\!\!\begin{array}{c} 0\\ 1\\ 0 \end{array}\!\!\right) + c_2 \ell \left(\!\!\begin{array}{c} m-1\\ 0\\ m \end{array}\!\!\right),\quad
\vb{\~v_3} = \left(\!\!\begin{array}{c} 0\\ 0\\ 1 \end{array}\!\!\right) + c_2 \ell \left(\!\!\begin{array}{c} m-1\\ 1\\ 0 \end{array}\!\!\right),
\nonumber\\
\eea
up to $O(\ell^2)$ terms.

Next we compute the deformed Riemann invariants.
It is non-trivial to find the correct integrating factor for the deformed Riemann invariants. 
To circumvent this issue, we take advantage of the fact that the total differential of each Riemann invariant is zero 
along the associated characteristic curve.
We expand the deformed Riemann invariants $\~R_1,\~R_2,\~R_3$ as
\eq{
\~R_j = r_j + \ell R^{(2)}_j(\@r_3) + O(\ell^2)\,,
\qquad j=1,2,3\,.
}
We have 
\eq{
\d \~R_j =0
\label{e:dRtildej=0}
}
along the characteristic curve $\d x/\d t = \~V_j$, for $j=1,2,3$. 
Expanding~\eqref{e:dRtildej=0} yields
\eq{
\d \~R_j = \grad_{\@r} \~R_j \cdot \d \@r=\grad_{\@r} \~R_j \cdot \bigg(\pdv{\@r_3}{t}\d t + \pdv{\@r_3}{x}\d x \bigg) = 0\,,
\qquad
j = 1,2,3\,,
\label{e:dRtildej=0_2}
}
where $\bfnabla_{\@r} = (\partial_{r_1},\partial_{r_2},\partial_{r_3})^T$.
Next we use \eqref{e:xt_3comp} and \eqref{e:KdVdeformed_eval} to rewrite \eqref{e:dRtildej=0_2} as
\eq{
(\grad_{\@r} \~R_j)^T (\~V_j \I_3 - A_3 ) \frac{\partial
\@r_3}{\partial x}\,\d t = 0\,,
\qquad 
j = 1,2,3\,,
}
along the curve $\d x/\d t = \~V_j$.
If the above differential must be zero for all $\frac{\partial
\@r_3}{\partial x}$, one can constrain each component to be zero, i.e.,
\eq{
(\grad_{\@r} \~R_j)^T ( \~V_j \I_3 - A_3 ) = \@0\,,\qquad 
j = 1,2,3\,.
\label{e:deformedinvariants_PDEsystem}
}
One can check that $\det(\~V_j \I_3 - A_3) = O(\ell^2)$ for $j = 1,2,3$, which allows for a nontrivial solution at $O(\ell)$.
For each $j=1,2,3$,
\eqref{e:deformedinvariants_PDEsystem} 
yields a system of three differential equations for $R^{(2)}_j$. 
Note that, 
even though it might not seem obvious a priori, these differential equations must necessarily be compatible
since we know that the system is integrable and therefore admits Riemann invariants.

We present the calculations for $R^{(2)}_1$ in detail.
Keeping terms up to $O(\ell)$, 
the first equation in~\eqref{e:deformedinvariants_PDEsystem} is trivially satisfied,
while the remaining two equations are 
\bse
\label{e:ds1}
\bea
\pdv{R^{(2)}_1}{r_2} & = \frac{(\E - (1-m)\K )^2}{3 (1-m)(r_2-r_1)^2}\,,
\label{e:ds1dr2}\\
\pdv{R^{(2)}_1}{r_3} & =  - \frac{ m^2 \E^2}{3(1-m)(r_2-r_1)^2}\,,
\label{e:ds1dr3}
\eea
\ese
and one can check that the equality of the mixed second derivatives, 
namely
$\partial^2 R^{(2)}_1/\partial{r_2}\partial{r_3} = \partial^2{R^{(2)}_1}\partial{r_3}\partial{r_2}$,
is indeed satisfied.
Next, we need to integrate~\eqref{e:ds1} to find $R^{(2)}_1$. 
We can integrate the equations manually, employing a process akin to
that of finding a potential for a conservative vector field.
%
We begin with~\eqref{e:ds1dr3} since it is simpler.
Because of the presence of elliptic integrals, it is convenient to
perform a change of variables from $r_1,r_2,r_3$ to $r_1,r_2$ and $m$. 
Solving~\eqref{e:m_def} for $r_3$ as a function of~$m$, we have
\eq{
\pdv{R^{(2)}_1}{m} = \pdv{R^{(2)}_1}{r_3}\pdv{r_3}{m} = \frac{\E^2}{3(1-m)(r_2-r_1)}\,.
}
Integrating this equation (with $r_1$ and $r_2$ held constant) then yields $R^{(2)}_1$ as
\bse
\bea
R^{(2)}_1 = \frac{g(m)}{3(r_2-r_1)}\,,
\label{e:s1m}
\\
\noalign{\noindent with}
g'(m) = \frac{\E^2}{1-m}\,.
\label{e:dgdm}
\eea
Note that we have taken the arbitrary function of $r_1$ and $r_2$ in
\eqref{e:s1m} to be zero.  By substituting \eqref{e:s1m} into
\eqref{e:ds1dr2} yields 
\eq{
g(m) = - (1-m)\K^2 + 2\E\K - \E^2\,,
\label{e:g}
}
\ese
and one can confirm that~\eqref{e:g} is indeed compatible with~\eqref{e:dgdm},
which means we have successfully integrated~\eqref{e:ds1},
obtaining the first approximate deformed Riemann invariant of the XT system as
\bse
\be
\~R_1 = r_1  + \frac{\ell}{3(r_2-r_1)}\qty(2\K\E - \E^2 -(1-m)\K^2) + O(\ell^2)\,.
\ee
One can apply an identical process to find the remaining deformed Riemann invariants as
\bea
\~R_2
= r_2 + \frac{\ell}{4\K}\bigg( 2\K\E +\frac{ \E^2}{1-m} -(1-m)\K^2 \bigg) + O(\ell^2) \,,
\\
\~R_3  = r_3 - \frac{\ell}{3(r_2-r_1)}\frac{m((1-m)\K^2-\E^2 )}{1-m} + O(\ell^2) \,.
\eea
\ese
The expressions of the deformed speeds and deformed Riemann invariants
may prove to be useful when
investigating the dynamics of weakly slanted wave fronts in the KP equation.

\begin{figure}[t!]
\centerline{\includegraphics[scale=0.90]{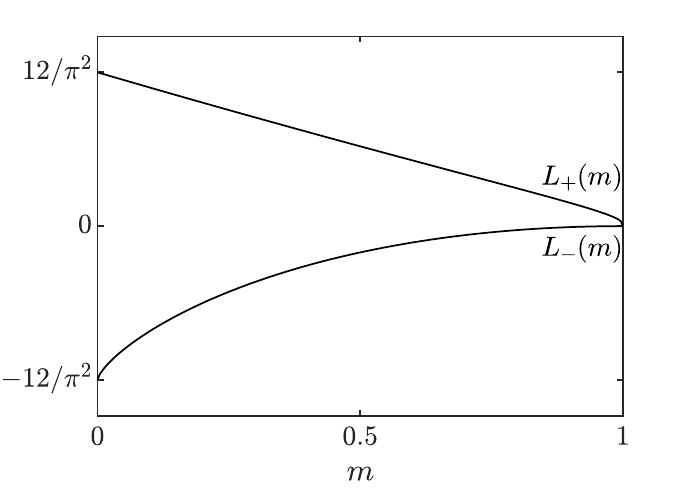}}
\kern-\bigskipamount
\caption{The functions $L_\pm(m)$ that determine the critical values
  $m = m_\pm$ at which two characteristic velocities of the XT system
  coalesce.}
\label{f:Lpm}
\end{figure}

\subsection{Hyperbolicity}
\label{sec:non-strict-hyperb}

The hyperbolicity of the three-component reduction~\eqref{e:xt_3comp} of the full KPWS 
can be determined by analyzing the eigenvalues of the coefficient matrix $A_3(\@r_3)$, given in~\eqref{e:xt_A3x3},
which are the characteristic
velocities $\tilde{V}_j$, $j=1,2,3$.  Because the characteristic
polynomial
\be
p(\lambda) = \det(A_3-\lambda \I_3) = -\lambda^3 + b_2 \lambda^2 + b_1 \lambda + b_0,
\label{e:KdVdeformed_charpoly}
\ee 
is a cubic with real coefficients, 
it has either three real roots or one real root and a complex conjugate pair.
Equation \eqref{e:KdVdeformed_evalapprox} demonstrates that the
$\tilde{V}_j$ are real for all $r_j \in \Real$ with $r_1 \le r_2 \le r_3$
when the deformation parameter $\ell = \sigma l_0^2$ is sufficiently small in magnitude.  
Because the coefficients $b_0,b_1,b_2$ in~\eqref{e:KdVdeformed_charpoly} 
are smooth functions of $\@r_3$, 
a bifurcation from all real roots to a complex conjugate pair can only occur if the discriminant of
$p(\lambda)$, 
\be 
D(\@r_3) = 4 b_1^3+b_2^2 b_1^2-18 b_0 b_2 b_1-4 b_0 b_2^3-27 b_0^2\,,
\label{e:KdVdeformed_discriminant}
\ee 
is zero.
To evaluate $D(\@r_3)$, we first simplify the calculation by restricting ourselves to
the set
\be
S = \left \{ \@r_3 = (0,m,1)^T ~ | ~ 0 \le m \le 1 \right \}\,,
\label{e:restricted_hyperbolicity}
\ee
and we simply write $D = D(m)$, which 
can be shown to be a quintic polynomial in $\ell$ with
coefficients depending on $m$.  
Setting $D(m) = 0$ and solving for~$\ell$, 
one finds two complex conjugate solutions, which are not of interest, plus 
two real solutions $\ell = L_-(m)$ and $\ell = L_+(m)$, the latter of which is a double root.
Explicitly,
\bea
L_+(m) = \frac{((2-m) E_m-2 (1-m) K_m) ((1+m) E_m - (1-m) K_m) ((1-m)
  K_m-(1-2 m) E_m)}{3 E_m (K_m-E_m) (E_m-(1-m) K_m)
  \left( 2 (2-m) K_m E_m - 3 E_m^2 - (1-m) K_m^2\right)},
\nonumber\\
\eea
while the expression for $L_-(m)$ is more complicated, so it is omitted for brevity.
The expansions of $L_\pm(m)$ for small~$m$ are 
\bea
L_+(m) &= \frac{12}{\pi^2}(1-m) + O(m^2), \\
L_-(m) &= -\frac{12}{\pi^2} \left (1 - \frac{4}{3} (2m)^{2/3} - m
\right ) + O(m^{4/3}).  \eea

Figure \ref{f:Lpm} shows that $L_+(m) > 0$ and $L_-(m) < 0$ for~$m \in [0,1)$.  
Therefore, there are two critical values of $m$:
$\ell = \pm l_0^2 = L_\pm(m_\pm)$.  
The fact that $\tilde{V}_j \in \Real$ for $|\ell|$ sufficiently small
implies that $D(m) > 0$ when $0 < m < m_\pm$, namely the XT system is
(strictly) hyperbolic when $m \in (0,m_\pm)$. 
When $m = m_\pm$, two characteristic velocities coalesce.  

In the case of the plus sign, 
the fact that $\ell = L_+(m_+)$ is a double root of $D(m) = 0$ implies that 
$\d D/\d m\,|_{m_+} = 0$.  
Then, in a neighborhood of $m = m_+$,
the discriminant \eqref{e:KdVdeformed_discriminant} exhibits parabolic behavior
\bea
D(m) = \frac{1}{2} \left . \frac{\mathrm{d}^2 D}{\mathrm{d} m^2}
\right |_{m_+} ( m - m_+ )^2 + O(m - m_+)^3\,, 
\eea
and it must be the case that $D(m)\ge0$, i.e.,  
$\d^2 D/\d m^2|_{m_+} > 0$, because $D(m) > 0$ for $0 < m < m_+$.   
Since
$m_+$ is the only point at which $D(m) = 0$, 
this implies $D(m) > 0$ for $m \in [0,m_+)\cup(m_+,1)$ and that 
the characteristic speeds are always real.  
Indeed, a direct calculation shows
that, when $m = m_+$,
\bse
\bea
\tilde{V}_1 &= \tilde{V}_2 = \frac{6 \left((1+m) E(m)^2-(1-m)^2
K(m)^2-2 (1-m) m K(m) E(m)\right)}{(1-m) K(m)^2 - 2
   (2-m) K(m) E(m)+3 E(m)^2}, \\
   \tilde{V}_3 &= \frac{2}{3} \left(3
   (m+1) + \frac{2 m E(m)}{(m-1)
   K(m)+E(m)}+\frac{2 m E(m)}{K(m)-E(m)} -\frac{2 (1-m)
   K(m)}{E(m)}\right) 
\eea
\ese
and there are three corresponding linearly independent eigenvectors
${\bf \tilde{v}_j}$, $j = 1,2,3$.  Consequently, we conclude
that the XT system is hyperbolic for $m \in (0,1)$ and strictly
hyperbolic when $m \ne m_+$.

In the case of the minus sign, the critical point $m_-$ satisfying 
$\ell = -l_0^2 = L_-(m_-)$ is a simple root of $D(m)$. 
Since $D(m) > 0$ for $0 < m < m_-$ and $D$ depends upon $m$ smoothly, 
it necessarily is the case that $\d D/\d m\,|_{m_-} < 0$, 
so that the discriminant \eqref{e:KdVdeformed_discriminant}
becomes negative in a right neigborhood of $m = m_-$.
This implies that, for $m_- < m < 1$, the XT system 
exhibits a complex conjugate pair of characteristic speeds and is not hyperbolic.

The above discussion of hyperbolicity of the XT system was limited to
the set $S$ defined in~\eqref{e:restricted_hyperbolicity}, 
where it was observed that a bifurcation occurs at the point $m = m_\sigma$.
However, using the scaling symmetry $r_j(x,t) \to a^2 r_j(ax,a^3t)$,
$q(x,t) \to a q(ax,a^3t)$ with $a = (r_3-r_1)^{-1/2}$ and the Galilean
symmetry $r_j(x,t) \to b + r_j(x-6bt,t)$, $q(x,t) \to q(x-6bt,t)$ 
with $b = -r_1$ \cite{Biondini_2017}, 
we can map any vector $\@r_3 = (r_1,r_2,r_3)^T \in \Real^3$ 
with $r_1 \le r_2 \le r_3$ 
to a vector $\~{\@r_3} = (0,m,1)^T\in S$.
For $\@r_3$, the bifurcation occurs on the surface
\bea
\Sigma_\pm = \left \{ 
  \@r_3 = (r_1,r_2,r_3)^T ~ \Big | ~ 
  \pm l_0^2 = L_\pm(m_\pm)(r_3 - r_1)^2,~~ 
  m_\pm = \frac{r_2 - r_1}{r_3 - r_1} \right \} .
\eea
In the case of the plus sign, the XT system is hyperbolic, and strictly so for $\@r_3 \notin \Sigma_+$.  
In the case of the minus sign, the XT system is hyperbolic so long as $(r_2 - r_1)/(r_3-r_1) < m_-$ where, 
$L_-(m_-) = -l_0^2(r_3 - r_1)^2$.  
When $(r_2 - r_1)/(r_3-r_1) > m_-$, the XT system loses hyperbolicity.

\section{The XY system}
\label{s:xy} 
\subsection{KPWS in a comoving frame and the XY system}

The third and final class of reductions of the KPWS we consider is that of time-independent solutions, to be defined precisely below.
While in Sections~\ref{s:yt} and \ref{s:xt} we considered reductions
that are evolutionary, exhibiting well-posed initial value problems
(at least when $\sigma = 1$), here we are considering a spatial
problem, independent of $t$, for the modulations.  As we will see,
this does not preclude dynamics in the full solution to KP itself.
In the previous sections we saw that, in order to ensure the compatibility of the XT and YT reductions 
with the KP equation, 
one must make sure that all three conservation of waves equations~\eqref{e:waveconservation} are satisfied.
We will see that this is also the case with stationary reductions of the KPWS.

Note that, even though one may think that a more general scenario is obtained by looking for traveling wave solutions, i.e., 
solutions that are stationary in a traveling frame of reference $(\~x,\~y,\~t)$, with $\~x = x - ct$, $\~y = y-dt$ and $\~t = t$,
this is not the case in practice.
This is because the Galilean and pseudo-rotation invariance of the KP~equation allow one to 
perform appropriate transformations of the dependent and independent variables to rewrite any traveling wave solution of the KP equation
as stationary in a suitable reference frame.
Since the KPWS preserves these invariances, the same transformations will also work for the KPWS, see Appendix~\ref{a:traveling} for details.

Based on the above discussion, consider situations in which the temporal derivatives in the original KPWS~\eqref{e:KPWS5}
can be neglected, which then yields
\be
A_5\,\partialderiv{\@r}{x} + B_5\,\partialderiv{\@r}{y} = 0\,.
\label{e:KPWS_XY}
\ee
Contrary to the reductions discussed in~\ref{s:yt} and~\ref{s:xt}, 
here the independence from one of the coordinates does not automatically result in 
a reduction in the number of degrees of freedom.
That is, all five dependent variables appear in~\eqref{e:KPWS_XY}.
Assuming invertibility of $A_5$ and $B_5$, one could equivalently 
write~\eqref{e:KPWS_XY} as an evolutionary system with respect to either $x$ or $y$, 
e.g., as
$\@r_x + C \,\@r_y = 0$. 
However, 
the resulting coefficient matrix $C = (A_5)^{-1} B_5$ is quite
complicated, and therefore the resulting system is difficult to analyze.

\subsection{Harmonic and soliton limits of the XY system}

Similar to the XT and YT systems, the XY system admits finite harmonic and soliton limits,
in which case the system simplifies considerably.
Specifically, in the harmonic limit ($r_2\to r_1^+$, corresponding to $m\to0$), 
the PDEs for $r_1$ and $r_2$ coincide, and \eqref{e:KPWS_XY} reduces to a four-component system of PDEs
for the vector $\@r' = (r_1,r_3,q,p)^T$, in which the coefficient matrices $A_5$ and $B_5$ are replaced by 
\bse
\bea
A_{4,o} = 
		\left(\begin{array}{ccccc}
			12 r_1-6 r_3 - \sigma q^2 & 0 & -q r_3 \sigma  & -q \sigma  \\
			0 & 6 r_3- \sigma q^2  & -q r_3 \sigma  & -q \sigma  \\
			0 & -6q & 12r_1 - 6r_3 -\sigma q^2 & 0\\
			0 & q & r_3 & 1 
		\end{array}\right),
\\
B_{4,o} = 
			\left(\begin{array}{ccccc}
				2 q \sigma  & 0 & r_3 \sigma  & \sigma  \\
				0 & 2 q \sigma  & r_3 \sigma  & \sigma  \\
				0 & 6 & 2\sigma q & 0 \\
				0 & -1 & 0 & 0
			\end{array}\right).
\eea
\ese
Similarly, in the soliton limit ($r_2\to r_3^-$, corresponding to $m\to1$), 
the PDEs for $r_2$ and $r_3$ coincide, and one obtains a four-component system for $\@r' = (r_1,r_3,q,p)^T$, 
with the matrices $A_5'$ and $B_5'$ replaced by 
\bse
\bea
A_{4,1} =
		\left(\begin{array}{ccccc}
		 	6 r_1- \sigma q^2  & 0 & -q r_1 \sigma  & -q \sigma  \\
			0 & 2 r_1+4 r_3 -\sigma q^2 & \frac{1}{3} q \sigma  (r_1-4 r_3) & -q \sigma  \\
			-2q & -4q & 2r_1 + 4r_3 -\sigma q^2 & 0 \\
			q & 0 & r_1 & 1
		\end{array}\right),
\\
B_{4,1} = 
		\left(\begin{array}{ccccc}
			2 q \sigma & 0 & r_1 \sigma  & \sigma  \\
			0 & 2 q \sigma  & -\frac{1}{3} \sigma  (r_1-4 r_3) & \sigma  \\
			2 & 4 & 2\sigma q & 0 \\
			-1 & 0 & 0 & 0
		\end{array}\right).
\eea
\ese
These systems coincide with the time-independent reduction of the harmonic and soliton limits studied in \cite{Biondini_2017,Biondini_2020},
where it was also shown that these systems are integrable.

\subsection{Riemann invariant, reduction and integrability}

In light of what we learned by studying the YT and XT systems, we expect that, when considering solutions that are independent of~$t$, 
the frequency $\omega$ will be one of the Riemann invariants.  
Indeed, in this case the three compatibility conditions~\eqref{e:waveconservation} yield immediately $\omega_x = \omega_y = 0$.
We now show that this expectation is correct.
In this case, however, the complexity of the system makes it impractical to use the direct approach 
based on the use of the characteristic relations and left eigenvectors
that was
used in the previous sections.
We therefore use an alternative approach, based on calculating the
total differential of $\omega = \omega(\mathbf{r})$ as
\bse
\be
\d\omega = \bfnabla_{\@r}\omega\cdot\d\@r = \bfnabla_{\@r}\omega\cdot\bigg( \partialderiv{\@r}x\d x + \partialderiv{\@r}y \d y \bigg)\,, 
\ee
with $\bfnabla_{\@r} =
(\partial_{r_1},\partial_{r_2},\partial_{r_3},\partial_q, \partial_p)^T$. 
The evolution of $\omega$ as dictated by the system~\eqref{e:KPWS_XY} along the characteristic coordinates $d y/d x = \lambda$ is then 
\be
\d\omega = \bfnabla_{\@r}\omega\cdot\big( \lambda \I_5 - C \big)\,\partialderiv{\@r}y\,\d x\,,
\label{e:domega}
\ee
\ese
with $C = (A_5)^{-1}B_5$.
Computing $\bfnabla_{\@r}\omega$, substituting in~\eqref{e:domega} and setting $\d\omega = 0$ then yields a linear equation that 
determines the characteristic speed~$\lambda$
as
\be
\lambda = \frac{2\sigma q}{V_2 - \sigma q^2},
\ee
with $V_2$ given in~\eqref{e:Vdef},
which confirms that $\omega$ is indeed a Riemann invariant for the system~\eqref{e:KPWS_XY}.%

As per the above discussion, in order for the KPWS to be compatible,
we must enforce $\omega_x = \omega_y = 0$.  
Following the procedures of section~\ref{s:yt} and~\ref{s:xt}, we partially diagonalize the $XY$ system~\eqref{e:KPWS_XY}
by performing a dependent coordinate transformation so that $\omega$ is one of the new dependent variables.
We then solve the resulting PDE by taking $\omega\equiv{}$const and obtain a one-parameter family of reduced four-component systems, 
parametrized by the constant value of $\omega$.
Once again, however, the calculations are more involved than in the previous cases.

The complication is that the expression~\eqref{e:qV} for $\omega$ does not allow one to uniquely obtain any one of the dependent variables
in terms of the others (recall~\eqref{e:def_k}).
The best one can do is to solve for $q$, which entails a choice of sign:
\vspace*{-1ex}
\be
q = \pm\sqrt{\sigma\bigg(\frac\omega k -  V\bigg)}\,.
\label{e:qdefXYreduced}
\ee
Following the same methods as in the previous sections, one can then obtain a four-component hydrodynamic system of equations for 
$\@v = (r_1,r_2,r_3,p)^T$.
A single coefficient matrix of the form
\be
\partialderiv{\@v}{x} + C_4'\,\partialderiv{\@v}{y} = 0\,,
\ee
is quite complicated. However, the system \eqref{e:KPWS_XY} can be transformed, using the same methods, into the concise form
\eq{
	A'\pdv{\vb u}{x} + B'\pdv{\vb u}{y} = 0\,,
}
where $A' = AT$ and $B' = BT$ and $\vb{u} =
(r_1,r_2,r_3,\omega,p)^T$. Once the PDE for $\omega$ is disregarded,
since $\omega \equiv{}$const solves it, we arrive at the system
\be
	A_4 '\partialderiv{\@ p}{x} + B_4 '\partialderiv{\@ p}{y} = 0\,,
\label{e:KPWS_XY4x4}
\ee
where $\@ p = (r_1,r_2,r_3,p)^T$, the coefficient matrices are
{\small
\bea
\fl
A_4 ' =  \left( \begin{array}{cccc}
	q \left(\left(c_1+1\right) \nu _1+q^2 (-\sigma )+V_1\right) & -\left(c_1+c_2-1\right) \nu _1 q & \left(c_2+1\right) \nu _1 q & -q^2 \sigma  \\
	\left(c_1+1\right) \nu _2 q & -q \left(\left(c_1+c_2-1\right) \nu _2+q^2 \sigma -V_2\right) & \left(c_2+1\right) \nu _2 q & -q^2 \sigma  \\
	\left(c_1+1\right) \nu _3 q & -\left(c_1+c_2-1\right) \nu _3 q & q \left(\left(c_2+1\right) \nu _3+q^2 (-\sigma )+V_3\right) & -q^2 \sigma  \\
	-\sigma\left(c_1+1\right) \nu _5-\left((\alpha -1) q^2\right) & \sigma\left(c_1+c_2-1\right) \nu _5 & \alpha  q^2-\sigma\left(c_2+1\right) \nu _5
	& q
\end{array} \right)\,,
\nonumber\\
\eea
}
\noindent
and
\be
\fl
B_4 ' =\left( \begin{array}{cccc}
	2 \sigma q^2 -\left(c_1+1\right) \nu _1 & \left(c_1+c_2-1\right) \nu _1 & -\left(\left(c_2+1\right) \nu _1\right) & q \sigma  \\
	-\left(\left(c_1+1\right) \nu _2\right) & 2 \sigma q^2 + \left(c_1+c_2-1\right) \nu _2   & -\left(\left(c_2+1\right) \nu _2\right) & q \sigma  \\
	-\left(\left(c_1+1\right) \nu _3\right) & \left(c_1+c_2-1\right) \nu _3 & 2\sigma q^2  -\left(c_2+1\right) \nu _3 & q \sigma  \\
	(\alpha -1) q & 0 & -\alpha q & 0 
 \end{array} \right)\,,
\ee
with
\be
	c_1 = \frac{\omega(\E-\K)}{16 m k^3 \K^3} \qquad \textrm{and}\qquad  c_2 = \frac{\omega \E}{16 k^3 \K^3 (1-m)}\,.
\ee
Note that $q$ is present in the matrices for readability, however its
definition is given in \eqref{e:qdef} in terms of the constant
parameter $\omega$ and the riemann type variables $r_j$.
Furthermore, one can use computer algebra software to perform the Haantjies tensor test on the resulting system.
Doing so, we have verified that, as in the case of the XT and YT systems, the Haantjies tensor of the reduced XY system
does indeed vanish identically, suggesting that the latter system is integrable as well.
The reduced system~\eqref{e:KPWS_XY4x4} admits hyperbolic or elliptic
regimes depending on, for example, the sign of the 
argument of the square root in~\eqref{e:qdefXYreduced}.

The above reduced XY system possesses a finite harmonic limit, similarly to those in the previous sections. 
Specifically, in the limit $r_2\to r_1^+$, the PDEs for $r_1$ and $r_2$ coincide, and the four-component system~\eqref{e:KPWS_XY4x4} reduces to
a $3\times3$ system in the independent variables $x$ and $y$ for the three-component dependent variable $\@r_3 = (r_1,r_3,p)^T$, with coefficient matrices
\bse
\bea
\fl
A_3 =
\left(
\begin{array}{ccc}
 (h_1+16) r_1-\frac{3}{8} (3 h_1+8) r_3 & \frac{1}{4} (h_1+4) r_3 & -\sqrt{\sigma } h_2 \\
 r_3-\frac{h_1 r_3}{8} & (h_1+4) r_1-\frac{3 h_1 r_3}{4}+9 r_3 & - \sqrt{\sigma }h_2 \\
 \frac{(h_1-8) r_3}{8 \sqrt{\sigma } h_2} & \frac{3 (h_1-4)r_3-4(h_1+4) r_1}{4 \sqrt{\sigma } h_2} & 1 \\
\end{array}
\right),
\\
\fl
B_3 =
\frac{1}{h_2}\left(
\begin{array}{ccc}
 -\left(\sqrt{\sigma } (16 (h_1+4) r_1+(40-17 h_1) r_3)\right) & -2 (h_1+4) r_3 \sqrt{\sigma } & \sigma h_2 \\
 (h_1-8) r_3 \sqrt{\sigma } & -2 \sqrt{\sigma } (8 (h_1+4) r_1+(20-7 h_1) r_3) &  \sigma h_2 \\
 0 & -h_2 & 0 \\
\end{array}
\right),
\eea
\ese
where 
\be
h_1 = \frac{\pi \omega}{\sqrt{(r_3 - r_1)^3}},\qquad
h_2 = 8 \sqrt{(h_1 - 2) r_3 - (h_1 + 4) r_1}.
\ee
Like with the XT and YT reductions, however, the system~\eqref{e:KPWS_XY4x4} does not admit a finite soliton limit in general,
since $\omega\to0$ as $m\to1$ [cf.~\eqref{e:klomegadef}], 
which is incompatible with having $\omega=\mathrm{const}\ne0$ in~\eqref{e:KPWS_XY4x4}.

\subsection{Stationary solutions of the KP equation and Whitham modulation system for the Boussinesq equation}

Importantly, even though the system~\eqref{e:KPWS_XY} describes stationary solutions of the KPWS, 
the corresponding solutions of the KP equation are \textit{not} stationary, unless $\omega=0$.
On the other hand, if $\omega=0$, the modulated solutions of the KP equation described by~\eqref{e:KPWS_XY}
are also stationary.
This point is relevant because stationary solutions of the KP equation \eqref{e:KP} 
satisfy versions of the Boussinesq equation, namely \cite{AC1991,AS1981}
\be
u_{\tau\tau} - c^2 u_{xx} + \sigma (6uu_x + \epsilon^2 u_{xxx})_x = 0\,,
\label{e:Boussinesq}
\ee with $\tau=y$ and $c=0$.  The case $\sigma = +1$ is the ``good''
Boussinesq equation with real linear dispersion.  Boussinesq derived
the ``bad'' version ($\sigma = -1$) as a long-wavelength model of
water waves, whose linearized equation is ill-posed
\cite{boussinesq_theorie_1872}.  Therefore, the modulation system
\eqref{e:KPWS_XY4x4} with $\omega=0$ is also the genus-1 Whitham
modulation equations for the above Boussinesq equations.  This is
noteworthy because the Boussinesq equations~\eqref{e:Boussinesq} are
associated with a $3\times3$ Lax pair (e.g., see
\cite{AC1991,AS1981}), which significantly complicates the
analysis, and as a result, the development of Whitham modulation
theory via, e.g., finite gap integration, has not been formulated yet.

We point out that, when $\omega = 0$, the modulation system \eqref{e:KPWS_XY4x4} greatly
simplifies because $c_1 = c_2 = 0$ and $q = \pm \sqrt{-\sigma V}$.  
Moreover, when $\omega=0$, the system~\eqref{e:KPWS_XY4x4}
remains well-defined both in the harmonic and the soliton limits.

\section{On the compatibility and integrability of the full KPWS}
\label{s5remarks}

Recall that the first conservation of waves equation, namely~\eqref{e:waveconservation1}, 
is one of the equations that eventually yield~\eqref{e:KPWS-1} (as in the KdV equation),
while the second conservation of waves equation, namely~\eqref{e:waveconservation2}, 
yields the evolution equation for $q$, namely, \eqref{e:KPWS-2}.
However, we have seen that the original, five-component KPWS~\eqref{e:KPWS5}
is not automatically compatible with the third conservation of waves equation, namely \eqref{e:waveconservation3}.
In this section, we investigate the question of the compatibility and integrability of the full KPWS~\eqref{e:KPWS}.
Specifically, 
we show that
when fields are independent of $x$, $y$ or $t$ the full six-component KPWS~\eqref{e:KPWS} 
becomes compatible,
and one recovers the results of the previous sections.  
The calculations in this section also provide an alternative way to obtain those results.

\subsection{Compatibility and integrability of the full YT system}
\label{s:5.1}

We begin by studying the compatibility and integrability of the ``full YT system'', 
namely the reduction of the overdetermined, six-component KPWS~\eqref{e:KPWS} when all fields are independent of~$x$.
As mentioned in section~\ref{s:yt}, 
under the assumption that $k$, $l$ and $\omega$ do not depend on $x$, the closure conditions~\eqref{e:waveconservation} immediately imply that 
$k$ is independent of both $y$ and $t$. 
For clarity, let us set $k = k_0$, with $k_0$ a real positive constant. 
Then \eqref{e:qconstraint} is satisfied trivially.
Moreover, the relation $k = k_0$ provides an algebraic constraint among the variables $r_1$, $r_2$ and $r_3$, which implies that only two of them are independent. 
Writing the the resulting system of equations in term of the variables $s_2$ and $s_3$ defined in section~\ref{s:yt}, 
one then obtains~\eqref{e:KPWS_yt_s2s3q} together with 
\bse
\label{e:KPWS_s}
\bea
\partialderiv{r_1}{y} + \alpha \partialderiv{s_2}{y} = 0,
\label{e:KPWS_s_d} \\
\sigma \partialderiv{p}{y} +  \partialderiv{r_1}{t} + 2 \sigma q \partialderiv{r_1}{y} + \sigma \nu_1 \partialderiv{q}{y} = 0\,.
\label{e:KPWS_s_e}
\eea
\ese
Altogether, \eqref{e:KPWS_yt_s2s3q} and~\eqref{e:KPWS_s}
are a system of five equations for the dependent variables $(s_3,s_2,q,r_1,p)$, 
which is partially decoupled since the variables $r_1$ and $p$ do not appear in \eqref{e:KPWS_yt_s2s3q}.
Hence, the system can be solved for the variables $(s_3,s_2,q)$, 
and $r_1$ and $p$ obtained from~\eqref{e:KPWS_s_d} and~\eqref{e:KPWS_s_e} by direct integration. 
Hence, we just need to focus on equations~\eqref{e:KPWS_yt_s2s3q} subject to the constraint~\eqref{e:k_via_s2}.

Note, that, for fixed $k_0>0$ the algebraic
equation~\eqref{e:k_via_s2} gives a one-parameter family of functions
of the form $s_3 = s_3 (s_2; k_0)$.  (Here we chose to view $s_3$ as a
function of $s_2$, but the results are equivalent if we interchange
$s_3 \leftrightarrow s_2$.)  Observe that
\[
\partialderiv{s_3}{t} = \deriv{s_3}{s_2} \partialderiv{s_2}{t}, \qquad 
\partialderiv{s_3}{y} = \deriv{s_3}{s_2} \partialderiv{s_2}{y}.
\]
Substituting into~\eqref{e:KPWS_yt_s3_2} and using~\eqref{e:KPWS_yt_s2_2}, 
one can verify that~\eqref{e:KPWS_yt_s3_2} is identically satisfied, 
which allows us to further reduce the analysis of the system to the coupled 
equations~\eqref{e:KPWS_yt_s2_2} and~\eqref{e:KPWS_yt_q_2}. 
Before we proceed further with the analysis of these equations, 
note that the constraint~\eqref{e:k_via_s2} can be equivalently written as
\bse
\label{e:s2s3constraint}
\be
s_3 = 4 k_0^2 m K_m^2\,, 
\label{e:s3constraint}
\ee
where we used the relation $s_2 = s_3/m$ and the fact that $k_0$ is a positive constant. 
The advantage of \eqref{e:s3constraint} is that it also allows us to express $s_2$ as
\be
s_2 = 4 k_0^2 K_m^2\,.
\label{e:s2constraint}
\ee
\ese
Equations~\eqref{e:s2s3constraint} show that $m$ is in fact a ``natural" variable for parametrising both $s_3$ and~$s_2$. 
Therefore, we now aim to replace~\eqref{e:KPWS_yt_s2_2} with a corresponding equation containing $m$ and $q$, 
which is promptly achieved by noting that
\[
\partialderiv{m}{t} = \deriv m{s_2} \partialderiv{s_2}{t} \qquad 
\partialderiv{m}{y} = \deriv m{s_2} \partialderiv{s_2}{y},
\]
with
\[
\deriv{m}{s_2} = \frac{m (1-m)}{4 k_0 ^2 K_m (E_m - (1-m) K_m)}.
\]
Substituting the above expressions into~\eqref{e:KPWS_yt_s2_2} and~\eqref{e:KPWS_yt_q_2}, 
we obtain the two-component system
\bse
\label{e:KPWS_YT_2comp}
\bea
\partialderiv{m}{t} + 2 \sigma q \partialderiv{m}{y}  + \sigma \Phi_1(m) \partialderiv{q}{y} = 0, 
\label{e:KPWS_YT_2comp_a} \\
\partialderiv{q}{t} + 2 \sigma q \partialderiv{q}{y}  + k_0 ^2 \Phi_2(m) \partialderiv{m}{y} = 0,
\label{e:KPWS_YT_2comp_b} 
\eea
\ese
where
\bse
\bea
\Phi_1(m) = 
  \frac{2 m (1-m) K_m \left (3 E_m^2 - 2 (2-m) E_m K_m + (1-m) K_m^2 \right )}{3 E_m (E_m - K_m) \left (E_m - (1-m) K_m \right)}\,, 
\\
\Phi_2(m)  = 
  \frac{8 \left(3 E_m^2 - 2 (2-m) E_m K_m + (1-m) K_m^2\right)}{m (1-m)}\,.
\eea
\ese

The system~\eqref{e:KPWS_YT_2comp} explicitly contains the constant parameter $k_0$, 
which cannot be eliminated by a rescaling of the dependent and independent variables.
The system~\eqref{e:KPWS_YT_2comp}, which is integrable, can be brought into the diagonal form
\be
\partialderiv{R_\pm}{t} = \lambda_\pm  \partialderiv{R_\pm}{y}, \qquad i =1,2
\ee
where the characteristic speeds $\lambda_\pm$ 
(i.e., the eigenvalues of the $2 \times 2$ coefficient matrix associated with the system~\eqref{e:KPWS_YT_2comp}), 
which are the same as for~\eqref{e:ytcharacteristiccurves},
are now expressed as
\[
\lambda_\pm = 2 \sigma q \pm k_0  \sqrt{\sigma \Phi_1(m) \Phi_2(m)}\,,
\]
and the associated Riemann invariants, which also define the change of variables 
$(m,q) \mapsto (R_{+},R_{-})$, 
and which are the same as \eqref{e:yt_2x2Riemanninvariants}, are now expressed as
\[
R_\pm = q \pm k_0  \int_0^m \sqrt{\frac{\Phi_2(\mu)}{\sigma \Phi_1(\mu)}}\,\d\mu\,.
\]

It is well known that systems of the 
form~\eqref{e:KPWS_YT_2comp} are integrable by the hodograph method, 
and the general solution $(R_{+}(y,t),R_{-}(y,t))$ is given by
(locally, i.e., in a  neighbourhood of points where $\partial{R_\pm}/\partial{y} \neq 0$)
\[
x + \lambda_\pm (R_{+},R_{-}) t + w_\pm(R_{+},R_{-}) = 0,
\]
where $w_\pm((R_{+},R_{-}))$ are solutions of the following system of linear PDEs:
\[
\frac1{w_{+} - w_{-}}{\partialderiv{w_\pm}{R_{\mp}}} = \frac1{\lambda_{+} - \lambda_{-}}{\partialderiv{\lambda_\pm}{R_{\mp}}}\,.
\]

One can look for further reductions by looking for solutions such that
$m = m_0$, with $m_0$ a constant value in the interval $[0,1]$.  Under
this assumption, equation~\eqref{e:KPWS_YT_2comp_a} implies the
constraints $\Phi_1(m_0) = 0$.  The only solutions to this constraint
arise when $m_0 = 0$ and $m_0 = 1$, i.e., in the harmonic and soliton
limit, respectively.  However, the case $m_0 = 1$ where $r_2 = r_3$ in
the soliton limit needs to be treated separately, since the
system~\eqref{e:KPWS_YT_2comp} has been derived under the assumption
that $k_0 >0$ whereas the condition $r_2 = r_3$ implies that
$k = k_0 = 0$.

\subsection{Compatibility and integrability of the full XT system}

Next we consider the reduction of the six-component full KPWS~\eqref{e:KPWS} when fields are independent of~$y$.
Imposing that $k$, $l$ and $\omega$ are $y-$independent, the closure 
conditions~\eqref{e:waveconservation} immediately imply that $l$ is constant.
Hence, setting $l = l_0$, with $l_0$ a fixed real constant, 
we can write 
\vspace*{-1ex}
\be
\label{e:qdep}
q = \frac{l_0}{k} = \frac{2 l_0 K_m}{\sqrt{r_3 - r_1}},
\ee
implying that $q$ is functionally dependent on $r_1$, $r_2$ and $r_3$. 
The corresponding reduction of the full KPWS~\eqref{e:KPWS} then coincides with~\eqref{e:KPWS_xt}, 
which reads in component form as
\bse
\label{e:KPWSyind}
\bea
\label{e:KPWSyind1}
&\partialderiv{r_{i}}{t} + (V_{i} + \sigma q^2 - 2 \sigma q^2 ) \partialderiv{r_{i}}{x}  - \sigma \nu_{i} q \partialderiv{q}{x}  - \sigma q \partialderiv{p}{x} = 0 \qquad i =1,2,3  \\
\label{e:KPWSyind2}
&\partialderiv{q}{t} + (V_2 + \sigma q^2 - 2 \sigma q^2) \partialderiv{q}{x} - (4 - \nu_{4}) q \partialderiv{r_1}{x} - (2 + \nu_{4}) q \partialderiv{r_3}{x} = 0 \\
\label{e:KPWSyind3}
& \partialderiv{p}{x} + (1-\alpha) q \partialderiv{r_1}{x} + \alpha q \partialderiv{r_3}{x} + \nu_{5} \partialderiv{q}{x} = 0.
\eea
\ese
Expanding the derivatives in~\eqref{e:KPWSyind2} as 
\[
\partialderiv{q}{t} = \sum_{i=1}^{3} \partialderiv{q}{r_{i}} \partialderiv{r_{i}}{t}, \qquad \partialderiv{q}{x} = \sum_{i=1}^{3} \partialderiv{q}{r_{i}} \partialderiv{r_{i}}{x},
\]
where $q = q(r_1,r_2, r_3)$ is given by~\eqref{e:qdep}, and substituting the expressions for
$\partial{p}/\partial{x}$ and $\partial{r_{i}}/\partial{t}$ for  $i=1,2,3$ 
obtained from the remaining equations~\eqref{e:KPWSyind1} and~\eqref{e:KPWSyind3}, 
one can directly check that equation~\eqref{e:KPWSyind2} is identically satisfied. 
Therefore, the reduction~\eqref{e:KPWSyind} of the full KPWS reduces to 
equations~\eqref{e:KPWSyind1} and ~\eqref{e:KPWSyind3}, 
which are equivalent to the $3 \times 3$ diagonalisable sytem~\eqref{e:xt_3comp} 
plus the equation~\eqref{e:KPWSyind3}. 
This equation, given the solutions $r_1$, $r_2$ and $r_3$ of the system~\eqref{e:xt_3comp}, 
allows one to recover $p$ by direct integration.

\subsection{Compatibility and integrability of the full XY system}
\label{s:5.3}

Similarly to the previous reductions, 
if $k$, $l$ and $\omega$ do not depend on $t$, 
the closure conditions~\eqref{e:waveconservation} immediately imply that $\omega$ is constant.
Then we set $\omega = \omega_0$, where $\omega_0$ is a real constant. 
Hence, the definition of $\omega$~\eqref{e:def_omega} implies
\[
q^2 = \sigma \left(  \frac{\omega_0}{k} - V \right),
\]
i.e., $q= q(r_1,r_2,r_3)$ is a function of the variables $r_1$, $r_2$, $r_3$ only. 
Then, when all fields are independent of~$t$, the full six-component KPWS~\eqref{e:KPWS} reduces to
\eqref{e:KPWS_XY}, which, in component form, is
\bse
\label{e:tindep}
\bea
\label{e:tindep1}
\fl
(V_{i} - \sigma q^2) \partialderiv{r_{i}}{x} - q \nu_{i} \partialderiv{q}{x} - \sigma q \partialderiv{p}{x} + 2 \sigma q \partialderiv{r_{i}}{y} + \sigma \nu_{i} \partialderiv{q}{y} + \sigma \partialderiv{p}{y} = 0 \qquad i=1,2,3,\\
\label{e:tindep2}
\fl
(V_2 - \sigma q^2) \partialderiv{q}{x} - (4 - \nu_{4}) q \partialderiv{r_1}{x} - (2 +\nu_{4}) q \partialderiv{r_3}{x} + 2 \sigma q \partialderiv{q}{y}  + (4 - \nu_{4}) \partialderiv{r_1}{y} + (2 +\nu_{4})  \partialderiv{r_3}{y} = 0,\\
\label{e:tindep3}
\fl
\partialderiv{p}{x} + (1 - \alpha) q \partialderiv{r_1}{x} + \alpha q \partialderiv{r_3}{x} + \nu_{5} \partialderiv{q}{x} - (1-\alpha) \partialderiv{r_1}{y} - \alpha \partialderiv{r_3}{y},\\
\label{e:tindep4}
\fl
\partialderiv{k}{y}  - q \partialderiv{k}{x} - k \partialderiv{q}{x} = 0 \,.
\eea
\ese
Rearranging~\eqref{e:tindep1} and~\eqref{e:tindep3} with respect to the $y$-derivatives of $r_1$, $r_2$, $r_3$ and $p$, 
and substituting into the remaining equations, we verify that, under the assumptions above, both equations~\eqref{e:tindep2} 
and~\eqref{e:tindep4} are identically satisfied. 
Therefore, the system~\eqref{e:tindep} reduces to a diagonalizable system of four equations for the variables 
$r_1$, $r_2$, $r_3$ and $p$.

\section{Concluding remarks}
\label{s:conclusions}

In conclusion, in this work we investigated the two-dimensional reductions of the KPWS~\eqref{e:KPWS}
obtained when all fields are independent of one of the spatial or temporal coordinates.
We have also seen that, even though the reductions of the original five-component KPWS~\eqref{e:KPWS5}
are not integrable,
adding the sixth equation, namely~\eqref{e:qconstraint}, 
which enforces the compatibility with the conservation of waves,
results in an additional constant of motion, 
which not only makes the reductions of the full KPWS compatible, but it also makes each reduction integrable.

The fact that the original KPWS~\eqref{e:KPWS5} is not integrable might seem
surprising, since it is an asymptotic reduction of the KP equation, which is integrable.  
It is important to realize, however, that not all solutions of the KPWS~\eqref{e:KPWS5} 
describe modulated solutions of the KP equation.  
This is because not all solutions of the KPWS~\eqref{e:KPWS5}
automatically satisfy the third conservation of waves $k_y = l_x$.  
In other words, the original KPWS~\eqref{e:KPWS5} describes modulated one-phase solutions of the KP
equation only if its initial conditions are such that this condition
is satisfied at $t=0$ \cite{Biondini_2017}.

Turning to the full, six-component KPWS~\eqref{e:KPWS},
in general one does not expect an overdetermined quasi-linear system to be 
either compatible or integrable, 
so some mechanism of enforcing compatibility is required.  
In our previous work, we enforced the compatibility, and thereby
obtained integrable systems, by considering the harmonic or soliton limit, 
either of which results in a reduction in the number of modulation equations.  
In this work, we added to the catalog of integrable reductions of the KPWS 
by characterizing two-dimensional reductions of the KPWS.

The results of this work and the above discussion lead to the natural
issue of whether there are other integrable reductions of the KPWS,
and whether it is possible to identify all such integrable reductions.
In other words, the question is whether it is possible to identify suitable conditions 
that ensure that the full KPWS is compatible.  
We plan to investigate this question in future work.

We reiterate that, even though both the reduced YT, XT and XY systems
admit a finite harmonic limit, none of these systems admits a
well-defined soliton limit in general.  However, setting the constant
values of $k$, $l$, or $\omega$ for the YT, XT, or XY system,
respectively, to zero does result in well-defined soliton limits.

We should also mention that one could equivalently carry out all calculations by replacing the PDE for $q$ with the following 
simplified PDE,
as derived in \cite{Biondini_2017,ABR_2018}:
\[
\partialderiv qt + (V + \sigma q^2)\partialderiv qx + \Dy ( V + \sigma q^2) = 0\,.
\]
For brevity, however, we omit the details.

Finally, we reiterate that the XY reduction of the KPWS allowed us to explicitly obtain the Whitham
modulation system for the Boussinesq equation, which had not been
derived before.  It is hoped that this
novel system will prove to be as useful as the other reductions of the
KPWS.  

Another potential application of the results of this work are to situations 
in which initial or boundary data for the KPWS are chosen to be
independent of one independent variable.  In order to use the reduced
YT, XT, or XY systems, the soliton limit will not be available except
in specialized situations, namely when $k \equiv 0$, $l \equiv 0$, or
$\omega \equiv 0$.  Nevertheless, one interesting class of problems
are generalized Riemann problems consisting of abrupt transitions
between two periodic traveling waves.  The reduced KPWS obtained here
could be used to study certain generalized Riemann problems.

\bigskip
\section*{\normalsize Acknowledgments}

\kern-\medskipamount 
The authors thank the Isaac newton Institute for Mathematical Sciences for its support and hospitality 
during the program Dispersive Hydrodynamics when parts of the present work were undertaken.  
This work was supported by: EPSRC Grant Number EP/R014604/1.  
GB and AB were partially supported by the National Science Foundation under grant number DMS-2009487.  
MH was partially supported by NSF under the grant DMS-1816934.

\section*{Appendix}
\setcounter{section}1
\setcounter{subsection}0
\setcounter{equation}0
\def\thesection{\Alph{section}}
\def\theequation{\Alph{section}.\arabic{equation}}
\def\numparts{\refstepcounter{equation}%
     \setcounter{eqnval}{\value{equation}}%
     \setcounter{equation}{0}%
     \def\theequation{\Alph{section}.\arabic{eqnval}{\it\alph{equation}}}}
\def\endnumparts{\def\theequation{\Alph{section}.\arabic{equation}}%
     \setcounter{equation}{\value{eqnval}}}

\subsection{Coefficients matrices, harmonic and soliton limits, relations between elliptic integrals}

The coefficient matrices $A_5$ and $B_5$ of the KPWS~\eqref{e:KPWS5} are:
\bse
\label{e:A5x5&B5x5}
\bea
A_5 = \left( \begin{array}{ccccc}
  V_1 - \sigma q^2 & 0 & 0 & -\sigma \nu_1 q & - \sigma q \\
  0 & V_2 - \sigma q^2 & 0 & -\sigma \nu_2 q & - \sigma q \\
  0 & 0 & V_3 - \sigma q^2 & -\sigma \nu_3 q & - \sigma q \\
 - (4-\nu_4)q & 0 & -(2+\nu_4)q & V_2 - \sigma q^2 & 0 \\
  -(1-\alpha)q & 0 & \alpha q & 0 & 0
\end{array} \right)\,,
\label{e:A5def}
\\
B_5 = \left(\begin{array}{ccccc}
 2\sigma q&0&0&\sigma\nu_1&\sigma\\ 
0&2\sigma q&0&\sigma\nu_2&\sigma\\ 
0&0&2\sigma q&\sigma\nu_3&\sigma\\
4-\nu_4&0&2+\nu_4&2\sigma q&0\\
1-\alpha&0&\alpha&0&0
 \end{array}\right)\,.
\label{e:B5def}
\eea
\ese
The definitions of all the coefficients appearing in~\eqref{e:A5x5&B5x5} are given in \eqref{e:Vdef} through \eqref{e:nudef}. 

Next, for convenience, we list the limiting values of the coefficients appearing in 
the harmonic and soliton limits of the KP-Whitham system, since these coefficients appear in all reductions.
Recal that, in the harmonic limit, the elliptic parameter $m$ tends to $0$
and $r_2\mapsto r_1^+$. 
In this limit, the various coefficients then become
\bse
\bea
m = 0\,, \quad V = 4r_1+2r_3\,,\\
V_1 = V_2 = 12r_1-6r_3\,, \quad V_3 = 6r_3\,,\\
\nu_1 = \nu_2 = \nu_3 = \nu_5 = r_3,\quad 
\nu_4 = 4, \quad \alpha = 1\,.
\eea
\ese
Conversely, in the soliton limit the elliptic parameter tends to $1$
and $r_2\mapsto r_3^-$. 
The limiting value of the various coefficients in this case is
\bse
\label{a:soliton_coeff}
\bea
m=1\,, \quad V = r_1+2r_3\,,\\
V_1 = 6r_1\,, \quad V_2 = V_3 = 2r_1+4r_3\,,\\
\nu_1 = \nu_5 = r_1 \,,\quad 
\nu_2 = \nu_3 = \txtfrac{1}{3}(4r_3-r_1)\,,\quad 
\nu_4 = 2\,,\quad \alpha = 0\,.
\eea
\ese


In this work we use the elliptic parameter $m$ as opposed to the elliptic modulus $k$. 
Recall that the two are related as
$m=k^2$.
The complementary modulus is then simply 
${k'} ^2 = 1-k^2 = 1-m$.
While this choice is in line with modern works, 
it differs from the convention in~\cite{NIST} and its associated references. 
Thus the various ODEs from\cite{NIST} must be transformed accordingly.
Specifically, 
the derivatives of $K$ and $E$ with respect to the elliptic parameter $m$ are 
\bse
\bea
\dv{\K}{m} = \frac{\E-(1-m)\K}{2m(1-m)}\,, \label{e:dkdm_ODE}\\
\dv{\E}{m} = \frac{\E-\K}{2m}\,.\label{e:dedm_ODE}
\eea
\ese
In addition, we have 
\be
\dv[2]{\E}{m} = - \frac1{2m}\dv{\K}{m}\,. 
\label{e:d2e}
\ee

\subsection{Invariances, traveling wave and stationary solutions of the KP equation and KPWS}
\label{a:traveling}

Here we show how, using the invariances of the KP equation and the KPWS, one can map all traveling wave solutions 
of the KP equation and the KPWS (i.e., solutions that are stationary in a comoving reference frame) 
into solutions that are stationary in a slanted but fixed reference frame.

To begin, it is useful to consider how the KPWS~\eqref{e:KPWS5} with coefficient matrices $I_4$, $A_5$ and $B_5$
is affected by affine transformations of the independent variables.
Recall first that the KP equation~\eqref{e:KP} is invariant under Galilean boosts,
\bse
\label{e:KPgalilean}
\bea
u(x,y,t) \mapsto u'(x,y,t) = c + u(x',y,t)\,,
\\
v(x,y,t) \mapsto v'(x,y,t) = v(x',y,t)\,,
\eea
\ese
with $x' = x - 6ct$,
and ``pseudo-rotations'',
\bse
\label{e:KPpseudorotation}
\bea
u(x,y,t) \mapsto u'(x,y,t) = u(x',y',t)\,.
\\
v(x,y,t) \mapsto v'(x,y,t) = v(x',y',t) + a u(x',y',t)\,,
\eea
\ese
with $x' = x+a y-\sigma a^2 t$ and $y' =  y- 2 \sigma a t$,
and with $a$ and $c$ arbitrary real parameters.
Namely, if the $u(x,y,t)$ and $v(x,y,t)$ comprises any solution of the KP equation, so does the pair $u'(x,y,t)$ and $v'(x,y,t)$.
Also recall that the above transformations are mapped respectively into
\bse
\bea
\@r(x,y,t) \mapsto \@r'(x,y,t) = \@1_3c + \@r(x', y,t),\qquad
\\
q(x,y,t) \mapsto q'(x,y,t) = q(x', y,t),  
\\
p(x,y,t) \mapsto p'(x,y,t) = p(x', y,t) - c q(x', y,t)\,,
\\
\noalign{\noindent 
with $\@r_3 = (r_1,r_2,r_3)^T$, $\@1_3 = (1,1,1)^T$ and $x' = x -6 c t$, and}
\nonumber\\[-2ex]
\@r(x,y,t) \mapsto \@r'(x,y,t) = \@r(x',y',t),\quad 
\\
q(x,y,t) \mapsto q'(x,y,t) = a +  q(x',y',t),
\\
p(x,y,t) \mapsto p'(x,y,t) = p(x',y',t).
\eea
\ese
with $x' = x + a y - \sigma a^2 t$ and $y' = y - 2 \sigma a t$.
Finally, recall that both of these transformations leave the original KPWS~\eqref{e:KPWS5} invariant \cite{Biondini_2017}.
Namely, if $\@r(x,y,t)$, $q(x,y,t)$ and $p(x,y,t)$ are any solutions of the KPWS, so are $\@r'(x,y,t)$, $q'(x,y,t)$ 
and~$p'(x,y,t)$.

We now show that, using the above invariances, all one- and two-phase traveling wave solutions of the KP equation
can be transformed to a stationary reference frame.
These are the solutions of the KP equation that can be written in the form 
\bse
\label{e:utraveling2phase}
\bea
&u(x,y,t) = U(z_1,z_2)\,,
\\
&z_n = k_nx + l_n y - \omega_n t,\quad 
n=1,2\,.
\label{e:zphasedef}
\eea
\ese
We show below that this formulation includes both classes of non-resonant elastic two-soliton solutions, 
the genus-2 solutions,
as well as the Miles resonance solution, the one-soliton solutions and the genus-1 solutions as special cases.
Starting with the two-phase solution~\eqref{e:utraveling2phase}, we apply a pseudo-rotation and Galilean boost, to obtain the new solution
\bse
\bea
&u'(x,y,t) = c + U(z_1',z_2')\,,
\\
&z_n' = k_n x + l_n' y - \omega_n' t\,,\quad
l_n' = l_n - a k_n\,,\quad
\omega_n' = \omega_n + (6c+\sigma a^2)k_n + 2\sigma a l_n\,,\quad 
\eea
\ese
for $n=1,2$.
The new solution $u'(x,y,t)$ is obviously stationary if $\omega_1' = \omega_2' = 0$.
In turn, it is trivial to see that it is always possible to achieve $\omega_1'=\omega_2'=0$ by choosing
\bse
\label{a:ac}
\bea
&a = -\frac{k_2 \omega_1 - k_1 \omega_2}{2 \sigma(k_2 l_1 - k_1 l_2)}\,,
\\
&c = \frac{ 4\sigma\,(k_1 l_2 - k_2 l_1)( l_2\omega_1 - \,l_1\omega_2 ) + (k_2\omega_1 - k_1\omega_2)^2 }
   {24\sigma(k_2 l_1-k_1 l_2)^2}\,.
\eea
\ese
[Note that the denominators in~\eqref{a:ac} are always non-zero for genuine two-phase solutions.
Conversely, if $k_2l_1 = k_1l_1$ the expression $u(x,y,t) = U(z_1,z_2)$ describes a one-phase solution, 
in which case it is sufficient to simply apply a Galilean boost.] 

By definition, the two-phase representation~\eqref{e:utraveling2phase} obviously includes all the
genus-2 solutions of the KP equation (e.g., see~\cite{BBEIM1994}), 
of which the genus-1 solutions are a special case.
It should then be clear that both of the non-resonant elastic two-soliton solutions 
as well as the Miles resonance solution and the one-soliton solutions are also included
(since the former are obtained as a degeneration of the genus-2 solutions \cite{AbendaGrinevich2018a,AbendaGrinevich2018b},
and the latter are in turn a degeneration of the former \cite{PRL99p064103}).
Nonetheless, we can give a simple proof of this fact.
Recall that general soliton solutions of the KP equation can be obtained through the Wronskian formalism as
\cite{JMP47p033514}
\bse
\label{e:ugeneral}
\bea
u(x,y,t) = 6\, \frac{\partial^2}{\partial x^2}[\log\tau(x,y,t)]\,, 
\qquad
\tau(x,y,t) = \Wr(f_1,\dots,f_N)\,, 
\\
f_n(x,y,t) = \sum_{m=1}^M C_{n,m} \e^{\theta_m}\,,
\quad
\theta_m(x,y,t) =  K_m x + \sqrt3\,K_m^2 y -4K_m^3 t\,,\quad m=1,\dots,M\,.
\eea
\ese
In particular, the Miles resonance solution is obtained by taking $N=1$ and $M=3$, so that
$\tau(x,y,t) = \e^{\theta_1} + \e^{\theta_2} + \e^{\theta_3}$,
and the two classes of non-resonant elastic two-soliton solutions 
are obtained by taking $N=2$ and $M=4$ and the following:
(i) for the ``ordinary'' two soliton solutions, 
$f_1 = \e^{\theta_1} + \e^{\theta_2}$ and $f_2 = \e^{\theta_3} + \e^{\theta_4}$;
(ii) for the ``asymmetric'' two soliton solutions, 
$f_1 = \e^{\theta_1} - \e^{\theta_4}$ and $f_2 = \e^{\theta_2} + \e^{\theta_3}$.
The Miles resonance solution is then cast in the framework of~\eqref{e:utraveling2phase} by simply writing
$\tau(x,y,t) = \e^{\theta_1}\,( 1 + \e^{z_1} + \e^{z_2})$, with $z_1 =  \theta_2-\theta_1$ and $z_2 =  \theta_3-\theta_1$,
since the common factor $\e^{\theta_1}$ disappears from the solution 
(because the $\theta_j$ are linear in~$x$) \cite{JMP47p033514}.
Similarly, for the ordinary two-soliton solution we have
$\tau(x,y,t) = \e^{\theta_1+\theta_3} + \e^{\theta_1+\theta_4} + \e^{\theta_2+\theta_3} + \e^{\theta_2+\theta_4}
 = 2\e^{\frac12(\theta_1+\theta_2+\theta_2+\theta_4)}(\cosh z_1 + \cosh z_2)$, 
where $z_1 = \frac12(\theta_1 + \theta_3 - \theta_2 - \theta_4)$
and $z_2 = \frac12(\theta_2 + \theta_3 - \theta_1 - \theta_4)$, 
and a similar representation works for the asymmetric two-soliton solution.

Finally, to complete our proof, we now show that no solutions containing more than two independent phases 
can be traveling wave solutions of the KP equation.
(In fact, this statement applies to general nonlinear evolution equations in two spatial dimensions.)
To see this, consider a generic $N$-phase solution $u(x,y,t) = U(z_1,\dots,z_N)$, with 
$z_n$ still given by~\eqref{e:zphasedef} for $n=1,\dots,N$.
If $u(x,y,t)$ is a traveling wave solution, there exists a coordinate transformations
$(x,y,t)\mapsto(X,Y,T)$ with $X = x - ct$, $Y = y - dt$ and $T = t$, 
such that $u(x,y,t) = u'(X,Y)$.
But the transformation yields $z_n = k_n(X+ct)+l_n(Y+dt) - \omega_nt$, 
so in order for $u'(X,Y)$ to be independent of~$T$, 
we need $c$ and $d$ such that 
\be
k_nc+l_nd = \omega_n\,,\qquad n=1,\dots,N\,.
\label{e:cdsystem}
\ee
If $N=1$, there are an infinite number of solutions to~\eqref{e:cdsystem}.  
(In particular, one can set $d=0$ and take $c = \omega_1/k_1$.)
If $N=2$, \eqref{e:cdsystem} admits a unique solution, given by
$c = (\omega_1 l_2 - \omega_2 l_1)/(k_1 l_2 - k_2 l_1)$ and 
$d = - (\omega_1 k_2 - \omega_2 k_1)/(k_1 l_2 - k_2 l_1)$.
If $N>2$, however, the system~\eqref{e:cdsystem} is overdetermined, and no solution exists.
(Here we assume that all phases are truly independent, which implies $k_n l_{n'} - k_{n'} l_n \ne 0$ for all
$n,n'=1,\dots,N$ with $n\ne n'$.
If this condition is violated, one can express the same solution with a smaller number of independent phases.)

\subsection{Haantjes tensor test for integrability}
\label{a:Haantjes}

An efficient criterion to test the diagonalizabiliy for a hydrodynamic system
that does not require the computation of the eigenvalues and eigenvectors of the coefficient matrix
was outlined in \cite{MathAnn2007}, 
involving the vanishing of the Haantjes tensor associated with the coefficient matrix.
Specifically, for strictly hyperbolic systems, \cite{MathAnn2007} gives the the following theorem 
as a necessary condition for diagonalizability:
``A hydrodynamic type system with mutually distinct characteristic speeds is diagonalizable 
if and only if the corresponding Haantjes tensor is identically zero.''

The calculation of the Haantjes tensor requires calculation of the Nijenhuis tensor first. 
The Nijenhuis tensor of a matrix $M^i _j$ is defined as
\be
N ^i _{jk} = 
  M^p _j \partial_{u^p}M^i _k 
  - M^p _k \partial_{u^p}M^i _j 
  - M^i _p ( \partial_{u^j}M^p _k -\partial_{u^k}M^p _j)
\ee
where $\partial_{u^k}= \partial/\partial{u^k}$. 
In our case, the matrix $M^i _j$ is the corresponding coefficient matrix of the system for which diagonlizability 
is being tested.
Once the Nijenhuis tensor is known, the Haantjes tensor can be obtained as
\be
H^i _{jk} = N ^i _{pr}M^p _j M^r _k - N ^p _{jr}M^i _p M^r _k -N ^p _{rk}M^i _p M^r _j + N ^p _{jk}M^i _r M^r _p\,.
\ee
The calculation of the various tensors below as applied to the various systems discussed in this work 
was performed using the Mathematica software package.

\def\title#1{``#1''}
\let\reftitle=\title


\section*{\normalsize References}
\begin{thebibliography}{1}

\bibitem{AbendaGrinevich2018a}
S. Abenda and P. G. Grinevich,
``Rational degenerations of M-curves, totally positive Grassmannians and KP2-solitons'',
Commun. Math. Phys. {\bf361}, 1029--1081 (2018)

\bibitem{AbendaGrinevich2018b}
S. Abenda and P. G. Grinevich,
``Real soliton lattices of the Kadomtsev-Petviashvili II equation and desingularization of spectral curves: The Gr$^\mathrm{TP}$(2,4) case''
Proc. Steklov Inst. Math. {\bf302}, 7--22 (2018)

\bibitem{ABR_2018}
M.~J. Ablowitz, G. Biondini, and I. Rumanov, 
 ``Whitham modulation theory for (2+1)-dimensional equations of Kadomtsev-Petviashvili type,'' 
  J. Phys. A: {\bf51}, 215501 (2018)
  
\bibitem{Biondini_2017}
M.~J. Ablowitz, G. Biondini, and Q. Wang, 
 ``Whitham modulation theory for the Kadomtsev-Petviashvili equation,'' 
  Proc. Royal Soc. A {\bf 473,} 20160695 (2017)

\bibitem{AC1991}
M. J. Ablowitz and P. A. Clarkson, 
\textit{Solitons, nonlinear evolution equations and inverse scattering} 
(Cambridge University Press, 1991).

\bibitem{AS1981}
M. J. Ablowitz and H. Segur, 
\textit{Solitons and the inverse scattering transform} 
(SIAM, Philadelphia, 1981)

\bibitem{BWK16}
F. Baronio, S. Wabnitz, and Y. Kodama,
\title{Optical Kerr spatiotemporal dark-lump dynamics of hydrodynamic origin},
Phys. Rev. Lett. {\bf116}, 173901 (2016) 

\bibitem{BBEIM1994}
E. D. Belokolos, A. I. Bobenko, V. Z. Enol’skii, A. R. Its and V. B. Matveev, 
\textit{Algebro-geometric approach to nonlinear integrable equations} 
(Springer, Berlin, 1994)
  
\bibitem{PRL99p064103}
G. Biondini,
\reftitle{Line soliton interactions of the Kadomtsev-Petviashvili equation},
\journal{Phys. Rev. Lett.}{99}, 064103 (2007)

\bibitem{JMP47p033514}
G. Biondini and S. Chakravarty,
\reftitle{Soliton solutions of the {Kadomtsev-Petviashvili} {II} equation},
\journal{J. Math. Phys.}{47}, 033514 (2006).

\bibitem{Biondini_2020}
G. Biondini, M.~A. Hoefer, and A. Moro, 
 ``Integrability, exact reductions and special solutions of the {KP}-Whitham equations,'' 
\journal{Nonlinearity}{33}, 4114--4132 (2020)

\bibitem{IP17p937}
M. Boiti, F. Pempinelli, A.~K. Pogrebkov, and B. Prinari, 
\title{Towards an inverse scattering theory for non-decaying potentials of the heat equation}, 
\journal{Inv. Probl.}{17}, 937--957 (2001)

\bibitem{JMP44p3309}
M. Boiti, F. Pempinelli, A.~K. Pogrebkov, and B. Prinari, 
\title{Extended resolvent and inverse scattering with an application to KPI}, 
\journal{J. Math. Phys.}{44}, 3309--3340 (2003)

\bibitem{TMP159p721}
M.~Boiti, F.~Pempinelli, {A. K.} Pogrebkov, and B.~Prinari,
\reftitle{Building an extended resolvent of the heat operator via twisting transformations},
\journal{Theor. Math. Phys.}{159}, 721--733 (2009)

\bibitem{TMP165p1237}
M.~Boiti, F.~Pempinelli, {A. K.} Pogrebkov, and B.~Prinari,
\reftitle{On the equivalence of different approaches for generating multisoliton solutions of the {KPII} equations},
\journal{Theor. Math. Phys.}{165}, 1237--1255 (2010)

\bibitem{boussinesq_theorie_1872}
J.~V.~Boussinesq, \reftitle{Th\`eorie Des Ondes et Des Remous Qui Se
  Propagent Le Long d'un Canal Rectangulaire Horizontal, En
  Communiquant Au Liquide Contenu Dans Ce Canal Des Vitesses
  Sensiblement Pareilles de La Surface Au Fond},
\journal{J. Math. Pures Appl.}{17}, 55--108 (1872)

\bibitem{ElHoefer}
G. A. El and M. A. Hoefer,
\title{Dispersive shock waves and modulation theory},
{Phys. D \textbf{333}, 11--65 (2016)}

\bibitem{Ferapontov2006}
  E. V. Ferapontov and K. R. Khusnutdinova, ``The Haantjes tensor and
  double waves for multi-dimensional systems of hydrodynamic type: a
  necessary condition for integrability'', Proc. Roy. Soc. A {\bf
    462}, 1197--1219 (2006)

\bibitem{MathAnn2007}
E. V. Ferapontov and D. G. Marshall,
``Differential-geometric approach to the integrability of hydrodynamic chains: the Haantjes tensor'', 
Mathematische Annalen {\bf 339}, 61--99 (2005), 

\bibitem{hinch1991}
E. J. Hinch et al. 
\textit{Perturbation methods} 
(Cambridge University Press, 1991)

\bibitem{Hirota2004} 
R. Hirota,
\textit{The Direct Method in Soliton Theory}
(Cambridge University Press, 2004).

\bibitem{InfeldRowlands}
E. Infeld and G. Rowlands, 
\textit{Nonlinear waves, solitons and chaos} 
(Cambridge University Press, 2000)

\bibitem{KP1970}
B. B. Kadomtsev and V. I. Petviashvili,
\title{On the stability of solitary waves in weakly dispersive media},
Sov.\ Phys.\ Dokl. {\bf15} 539--541 (1970) 

\bibitem{Kodama2018}
Y. Kodama, 
\textit{Solitons in two-dimensional shallow water} 
(SIAM, 2018)

\bibitem{Konopelchenko}
B. Konopelchenko,
\textit{Solitons in multidimensions}
(World Scientific 1993)

\bibitem{Lonngren1998}
K. E. Lonngren, 
\title{Ion acoustic soliton experiments in a plasma},
Optical and Quantum Electronics {\bf30}, 615--630 (1998)

\bibitem{NMPZ1984}
S.~P. Novikov, S.~V. Manakov, L.~P. Pitaevskii, and V.~E. Zakharov, 
\textit{Theory of solitons. The inverse scattering method} 
(Plenum, New York, 1984)

\bibitem{NIST}
F. W. Olver, D. W. Lozier, R. F. Boisvert and C. W. Clark, 
\textit{NIST handbook of mathematical functions} 
(Cambridge, 2010)

\bibitem{PSK95} 
D. E. Pelinovsky, Y. A. Stepanyants, and Y. S. Kivshar,
\title{Self-focusing of plane dark solitons in nonlinear defocusing media},
Phys. Rev. E {\bf 51}, 5016--5026 (1995) 

\bibitem{NLTY21}
S. Ryskamp, M.~A. Hoefer, and G. Biondini, 
 ``Oblique interactions between solitons and mean flows in the Kadomtsev-Petviashvili equation,''
  Nonlinearity {\bf 34,} 3583--3617 (2021)

\bibitem{PRSA22}
S. Ryskamp, M.~A. Hoefer, and G. Biondini, 
``Modulation theory for soliton resonance and Mach reflection,''
  Proc. Roy. Soc. A {\bf 478,} 20210823 (2022)

\bibitem{JFM21}
S. Ryskamp, M.~D. Maiden, G. Biondini, and M.~A. Hoefer, 
 ``Evolution of truncated and bent gravity wave solitons: the mach expansion problem,''
  J. Fluid Mech. {\bf 909,} A24 (2021)

\bibitem{Smoller1994}
J. Smoller,
\textit{Shock waves and reaction diffusion equations}
(Springer, 1994)
  
\bibitem{TF85} 
S. K. Turitsyn and G. E. Fal'kovich,
\title{Stability of magnetoelastic solitons and self-focusing of sound in antiferromagnet},
Sov. Phys. JETP {\bf62}, 146--152  (1985)

\bibitem{Whitham1965}
G. B. Whitham,
``Non-linear dispersive waves,'' 
Proc. Roy. Soc. A {\bf283}, 238--261 (1965)

\bibitem{Whitham1974}
G. B. Whitham,
\textit{Linear and nonlinear waves}
(Wiley, 1974)

\bibitem{DerchyiWu2021}
D. Wu,
\title{The direct scattering problem for perturbed Kadomtsev-Petviashvili multi line solitons},
J. Math. Phys. {\bf62}, 091513 (2021)

\bibitem{DerchyiWu2022}
  D. Wu, \title{Stability of Kadomtsev-Petviashvili multi-line solitons},
arXiv:2205.07432 (2022)


\end{thebibliography}

\end{document}